\documentclass[
   pre,
   superscriptaddress,
   amsfonts,
   amssymb,
   amsmath,
   superscriptaddress,
   floatfix,
   eqsecnum,
      showpacs,
   preprintnumbers,
   byrevtex]{revtex4}
\usepackage{color}
\usepackage{graphicx}
\usepackage{morefloats}

\begin{document}

\newcommand{\vphi}{\varphi}
\newcommand{\bq}{\begin{equation}}
\newcommand{\be}{\begin{equation}}
\newcommand{\ba}{\begin{eqnarray}}
\newcommand{\eq}{\end{equation}}
\newcommand{\ee}{\end{equation}}
\newcommand{\ea}{\end{eqnarray}}
\newcommand{\tchi} {{\tilde \chi}}
\newcommand{\tA} {{\tilde A}}
\newcommand{\tq} {{\tilde q}}
\newcommand{\tphi} {{\tilde \phi}}
\newcommand{\tp} {{\tilde p}}
\newcommand{\tC} {{\tilde C}}
\newcommand{\sech} { {\rm sech}}
\newcommand{\pstar}{\mbox{$\psi^{\ast}$}}
\newcommand{\dn} { {\rm dn}}
\newcommand{\sn} { {\rm sn}}
\newcommand{\cn} { {\rm cn}}

\preprint{LA-UR 11-06383 : \today}


\newpage

\title{Forced Nonlinear Schr{\" o}dinger Equation with Arbitrary Nonlinearity}
\author{Fred Cooper}
\email{cooper@santafe.edu}
\affiliation{Santa Fe Institute, Santa Fe, NM 87501, USA}
\affiliation{Theoretical Division and Center for Nonlinear Studies, 
Los Alamos National Laboratory, Los Alamos, New Mexico 87545, USA}
\author{Avinash Khare} 
\email{khare@iiserpune.ac.in}
\affiliation{ Indian Institute of Science Education and Research, Pune 
411021, India}
\author{Niurka R. Quintero} 
\email{niurka@us.es} 
\affiliation{IMUS and Departamento de Fisica Aplicada I, E.U.P. Universidad de Sevilla, 41011 Sevilla, Spain}
\author{Franz G.  Mertens}
\email{Franz.Mertens@uni-bayreuth.de}
\affiliation{Physikalishes Institut, Universitat Bayreuth, D-95440 Bayreuth, Germany} 
\author{Avadh Saxena}
\email{avadh@lanl.gov}
\affiliation{Theoretical Division and Center for Nonlinear Studies, 
Los Alamos National Laboratory, Los Alamos, New Mexico 87545, USA}
\begin{abstract}
We consider the nonlinear Schr{\" o}dinger equation  (NLSE)    in 1+1 dimension with 
scalar-scalar  self interaction 
$\frac{ g^2}{\kappa+1} (  \psi^\star \psi)^{\kappa+1}$ in the presence of the external forcing terms of the form $r e^{-i(kx + \theta)} -\delta \psi$. We find new exact solutions for this problem and show that the solitary wave momentum is conserved in a moving  frame where $v_k=2 k$. These new exact solutions reduce to the constant phase solutions of the unforced problem when $r \rightarrow 0.$
 In particular we study the behavior of solitary wave solutions  in the presence of these external forces in a variational approximation which allows the position, momentum, width and phase of these waves to vary in time. We  show that the stationary solutions of the variational equations include  a solution close to the exact one and we study small oscillations around all  the stationary solutions. 
 We postulate that  the dynamical condition for instability is that 
$ dp(t)/d \dot{q} (t) < 0$,  where $p(t)$ is the normalized canonical momentum $p(t) = \frac{1}{M(t)} \frac {\partial L}{\partial {\dot q} }$, and $\dot{q}(t)$  is the solitary wave velocity. Here
$M(t) = \int dx \psi^\star(x,t)  \psi(x,t)$. Stability is also studied using a ``phase portrait" of the soliton, where its dynamics is represented by two-dimensional projections of its trajectory in the four-dimensional space of collective coordinates. 
 The criterion 
for stability of a soliton is that its trajectory is a closed single  curve with a positive sense of rotation around a fixed point. 
 We investigate the accuracy of our variational approximation and these criteria using numerical simulations of the NLSE.  We find that our criteria work quite well when the magnitude of the  forcing term is small compared to the amplitude  of the unforced solitary wave.  In this regime the variational approximation captures quite well the behavior of the solitary wave. 
\end{abstract}


\maketitle


\section{Introduction}
The nonlinear Schr{\" o}dinger equation (NLSE), with cubic and higher
nonlinearity counterparts, is ubiquitous in a variety of
physical contexts.  It has found several applications,
including in nonlinear optics where it describes pulse
propagation in double-doped optical fibers \cite{doped} and in Bragg
gratings \cite{Bragg}, and in Bose--Einstein condensates (BECs) where
it models condensates with two and three-body interactions \cite{bec1,bec2}. 
Higher order nonlinearities are found in the context of Bose
gases with hard core interactions \cite{hard}  and low-dimensional
BECs in which quintic nonlinearities
model three-body interactions \cite{three}.  In nonlinear optics a
cubic-quintic NLSE is used as a model for photonic
crystals \cite{crystals}. Therefore, it is important to ask the question
how will the behavior of these systems change if a forcing term
is also included in the NLSE.

The forced nonlinear Schr{\" o}dinger equation (FNLSE)  for   an interaction 
of the form $(\psi^\star \psi)^2$ has been recently studied \cite{mqb,Mpreprint}  using collective coordinate (CC) methods such as  time-dependent variational methods and the generalized traveling wave method (GTWM) \cite{qmb}.  In \cite{mqb} and  \cite{Mpreprint} approximate stationary solutions to the variational solution  were found and a criterion for the stability of these solutions under small perturbations was developed and compared to numerical simulations of the FNLSE.    Here we will generalize our previous study to arbitrary nonlinearity $(\psi^\star \psi)^{\kappa+1}$, with a special emphasis on the case $\kappa = 1/2$.
That is, the form of the FNLSE  we will consider is
\bq
 i \frac{\partial}{\partial t} \psi + \frac{\partial^2}{\partial x^2} \psi + {g }(\psi^\star \psi)^{ \kappa} \psi+\delta \psi  = r e^{-i( k x+ \theta)}.  \label{FNLSE}
\eq
The parameter $r$ corresponds to a plane wave driving term.  The parameter $\delta$ arises in discrete versions of the NLSE used to model discrete solitons in optical wave guide arrays and is a cavity detuning parameter \cite{cavity}.  We will find that having $ \delta <0$ allows for constant phase solutions of the CC equations. The externally driven NLSE arises in many physical situations such as charge density waves \cite{charge}, 
long Josephson junctions \cite{Josephson}, optical fibers
 \cite{optical} and plasmas driven by rf fields \cite{rf}.  What we would like to demonstrate here is that the stability criterion for the FNLSE solitons found for $\kappa = 1$ works for arbitrary $\kappa  \leq 2$, and that the collective coordinate method works well in predicting the behavior of the solitary waves when the forcing parameter $r$ is small compared to amplitude of the unforced solitary wave.   

The paper is organized as follows.  In Sec. II we show that in a 
comoving frame where $y=x+2kt$, the total momentum of the solitary wave 
$P_v$ as well as the  energy of the solitary wave $E_v$  is conserved. 
 In Sec. III we  review the exact solitary waves for $r=0$ and 
$\kappa$ arbitrary.  We show using Derrick's theorem \cite{ref:derrick} that these solutions are unstable for $\kappa > 2$ and arbitrary $\delta$, which we 
later verify in our numerical simulations. 
In Sec. IV we find  exact solutions  to the forced problem for $r \neq 0$ and find both plane wave as well as solitary wave and periodic  solutions for arbitrary $\kappa$.  We focus 
mostly on the case $\kappa=1/2$.  We find both finite energy density as well as finite energy solutions.
In section V we discuss the collective coordinate approach in the Laboratory frame. We will use the form of  the exact solution for the unforced problem with time dependent coefficients as the variational ansatz  for the  traveling wave for $\kappa <2$.  This is a particular example of a collective coordinate (CC) 
approach \cite{mqb,HF,stable1,cooper3}.  We will assume the forcing term is of the form  $r e^{-i(kx + \theta)} -\delta \psi.$  We will choose four collective coordinates, 
the width parameter $\beta(t)$, the position $q(t)$
 and momentum $p(t)$,  as well as  the phase $\phi(t)$ of the solitary wave. These CCs are related by the conservation of momentum in the comoving frame.  We will derive the effective Lagrangian for the collective coordinates and determine the equations of motion for arbitrary nonlinearity parameter $\kappa$.  In Sec. VI we show that the equations of motion for the collective coordinates simplify in the comoving frame.  For arbitrary $\kappa$ we determine the equations of motion for the collective coordinates, the stationary solutions
 ($\beta={\dot q} =\phi={\rm constant} $)  as well as the linear stability 
of these stationary solutions. We then specialize to the case $\kappa = 1/2$, 
and discuss the linear stability of the stationary solutions.
 The real or complex solutions to the small oscillation problem give a local indication of stability of these stationary solutions. This analysis will be confirmed in our numerical simulations.  For $\kappa = 1/2$ and $r \ll A$, where $A$ is the amplitude of the solitary wave, we find in general three stationary solutions. Two are near the solutions for $r=0$, one being stable and another unstable and one is of much smaller amplitude but turns out to be a stable solitary wave.

A more general question of stability for initial conditions having an arbitrary 
value of $\beta(t=0) \equiv \beta_0$ is provided by the phase portrait of the system found by plotting the trajectories of 
the imaginary vs. the real part of the variational wave function starting with an initial value of $
\beta_0$.  These trajectories are closed orbits as shown in  \cite{Mpreprint}.  Stability is related to whether the orbits show a positive (stable)
or negative (unstable) sense of rotation or a mixture (unstable).  Another method for discussing stability is to use 
 the dynamical criterion used previously  \cite{Mpreprint}   
in the study of the case $\kappa = 1$, namely whether the  $p({\dot q})$ 
curve has a branch with negative slope. If this is true, this implies instability. 
Here  $p(t)$ is the normalized canonical momentum 
$p(t) = \frac{1}{M(t)} \frac {\partial L}{\partial {\dot q} }$,  
$M(t) = \int dx \psi^\star(x,t)  \psi(x,t)$ and ${\dot q}(t) = v(t) $ is the  velocity of the solitary 
wave. These two approaches (phase portrait and the slope of the  $p(v)$ curve)  give complementary approaches to understanding the behavior of the numerical solutions of the FNLSE.  In Sec. VII we discuss how damping modifies the equations of motion by including a dissipation function.  We find that damping only effects the equation of motion for $\beta$ among the CC equations.  This damping allows the numerical simulations to find stable solitary wave solutions. 
In Sec. VIII we discuss our methodology for solving the FNLSE. We explain how we extract the parameters associated with the collective coordinates from our simulations.  We first show that our simulations reproduce known results for the unforced problem as well as show that the exact solutions of the forced problem we found are metastable.  On the other hand the  linearly stable stationary solutions to the CC equations are close to exact numerical  solitary waves of the forced NLSE with time independent widths, only showing small oscillations about a constant value of $\beta$. 
 In the numerical solutions of both the PDEs and the CC equations we  establish two results.  First,  for small values of the forcing term $r$ the CC equations give an accurate representation of the behavior of the width, position and phase of the solitary wave determined by numerically solving the NLSE.  Secondly,  both the phase portrait and $p(v)$ curves allow us to predict  the stability or instability of  solitary waves that start initially as an approximate variational solitary wave of the form of the exact solution to the unforced problem.  In Section IX we state our main conclusions. 

\section{Forced Nonlinear  Schr{\" o}dinger equation (FNLSE)  }
The action for the FNLSE is given by
\bq
\int L dt = \int dt dx   \left[ i \psi^\star  \frac{\partial}{\partial t} \psi
 -[  \psi^\star_x   \psi_ x - \frac{g }{\kappa+1} (\psi^\star \psi)^{ \kappa +1} - \delta \psi^\star \psi+re^{i(kx+\theta)} \psi +r  e^{-i(kx+\theta)}  \psi^\star ] \right]  . \label{action}
\eq
As shown in \cite{mqb}  the energy
\bq E =  \int dx  \left[ \psi^\star_x   \psi_ x - \frac{g }{\kappa+1} (\psi^\star \psi)^{ \kappa +1} -\delta \psi^\star \psi +re^{i(kx+\theta)} \psi +r  e^{-i(kx+\theta)}  \psi^\star \right]
\eq
is conserved.
Varying the action, Eq. \eqref{action} leads to the equation of motion: 
\bq
 i \frac{\partial}{\partial t} \psi + \frac{\partial^2}{\partial x^2} \psi + {g }(\psi^\star \psi)^{ \kappa} \psi+\delta \psi  = r e^{-i( k x+ \theta)} .  \label{psieq}
\eq
We notice that the equation of motion is invariant under the joint 
transformation $r \rightarrow -r$, $\psi \rightarrow - \psi$, thus 
if $\psi(x,t,r)$ is a solution, then so is $- \psi(x, t, -r)$ a solution. 
Letting $\psi (x,t) = e^{-i(kx+\theta)} u(x,t) $ we obtain the equation

\bq
 i \left[ \frac{\partial u(x,t) }{\partial t}  -
2 k \frac{\partial u(x,t)}{\partial x}\right]  - (k^2-\delta) u(x,t)+  \frac{\partial^2 u(x,t) }{\partial x^2} + {g }(u^\star u)^{ \kappa} u = r . 
\eq
Changing variables from $x$ to $y$  where  $y=x + 2 k t=x+v_{k} t$ we have for 
$u(x,t) \rightarrow v(y,t)$
\bq
 i  \frac{\partial v(y,t) }{\partial t}- (k^2-\delta) v(y,t)+  \frac{\partial^2 v(y,t) }{\partial y^2} + {g }(v^\star v)^{ \kappa} v = r  .  \label{veq}
\eq
Note that with our conventions, the mass in the Schr{\" o}dinger equation 
obeys $2m=1$, or   $m=1/2$  so that $k = m v_{k} = v_{k}/2$. This equation in 
the moving frame can be derived from a related action
\bq 
S_v = \int dt dy   \left[ i v^\star  \frac{\partial}{\partial t} v  -[  v^\star_{y}   v_ {y}-\frac{g }{\kappa+1} (v^\star v)^{ \kappa +1}+(k^2 - \delta)v^\star v+rv +r  v^\star ]\right] . ~~ \label{Actionv}
\eq
Multiplying Eq. (\ref{veq}) on the left by $v^\star_y$ and adding the complex conjugate, we get an equation for the time evolution of the momentum density  in the moving frame: $\rho_v =  \frac {i}{2}(v  v^\star_y - v^\star v_y)$, namely
\bq
\frac{\partial \rho_v}
{\partial t} + \frac{\partial j(y,t)}{\partial y} = r  v^\star_{y} +  r  v_{y} . 
\eq
Here
\bq
 j(y,t) =\frac{i}{2} \left(v^\star v_t - v v^\star_t \right) + |v_{y}|^2 + (\delta - k^2) |v|^2 +\frac{g}{\kappa+1} |v|^{2 \kappa+2} . 
 \eq
Integrating over all space
we get 
\bq
\frac{ d}{dt} \int dy  \rho_v (y,t) = F[y=+ \infty]- F[y=-\infty] ,
\eq
where
\bq
F[y] = - j(y,t) + r v^\star(y,t) +  r  v(y,t).
\eq
If the value of the boundary term is the {\it same} (or zero)  at $y= \pm \infty$ then  
 we find that the momentum $P_v$  in the moving frame is conserved:
\bq
P_v = constant = \int dy  \frac {i}{2}(v  v^\star_y - v^\star v_y) . 
\eq 
Using the fact that $\psi = u e^{-i(kx+\theta)}$ and defining 
\bq \label{momentum}
P=  \int dx  \frac {i}{2}(\psi \psi ^\star_x - \psi^\star \psi_x) , 
\eq
we obtain the conservation law
\bq
P(t) +M(t) k = P_v = constant,
\eq
where  $M(t) = \int dx \psi^\star \psi$.
If we further define 
\bq
p(t) = \frac{P(t)}{M(t)} , 
\eq
then we get the relationship:
\bq
P_v = M(t) (p(t)+k) = constant. 
\label{pv}
\eq
This equation will be useful when we consider variational approximations for the solution.

The second action  Eq. (\ref{Actionv}) also leads to a conserved energy:
\bq
E_v = \int dx \left[  v^\star_{y}   v_ {y}-\frac{g }{\kappa+1} (v^\star v)^{ \kappa +1}+(k^2 - \delta)v^\star v+r  v +r  v^\star \right] .
\eq
Using the connection that $\psi(x,t) = e^{-i(kx+ \theta)} v(y, t)$ we find that $E$ and $E_v$ are related.  Using the fact that 
\bq \psi_x^* \psi_x = k^2 v^* v + v_y^*v_y + ik (v^* v_y - v v^*_y) , \eq we obtain
\bq
E = E_v - 2 k P_v .
\eq

\section{Exact Solitary Wave Solutions when $r=0$}

Before discussing the exact solutions to the forced NLSE,  let us review the  
exact solutions for $r=0$ since these will be used as our variational trial functions later in the paper.  We can  obtain the exact solitary wave solutions when $r=0$  as follows.  We let
\bq
\psi(x,t) = A ~ \sech ^\gamma \left [ \beta (x- vt ) \right]  e^{i \left[ p\left(x- vt \right) + \phi(t)\right]} . \label{exact}
\eq
Demanding we have a solution by matching powers of $\sech$ we find:
\bq
p= v/2\,, ~~ \gamma =1/\kappa\,,~~  A^{2 \kappa} 
= \frac{\beta^2 (\kappa+1)}{g \kappa^2}\,, ~~
\phi = (p^2+  \beta^2/\kappa^2+\delta) t + \phi_0\,.
\eq
It is useful to connect the amplitude $A$ to $\beta$ and the  mass $M$ of the solitary wave.
\bq
M= A^2 \int dx~ \psi^*(\beta x) \psi (\beta x) = \frac{A^2}{\beta} C_1^0; \quad 
C_1^0 = \int_{-\infty}^{+\infty}  dy~ \sech^{2/k} (y) =  \frac{\sqrt{\pi } \Gamma \left(\frac{1}{\kappa }\right)}{\Gamma \left(\frac{1}{2}+\frac{1}{\kappa }\right)} . 
\eq
One can show that the energy $E = \int_{-\infty}^{+\infty} dx~ H$ of the solitary waves is given by
\bq
E=\frac{\sqrt{\pi} \Gamma(\frac{1}{\kappa})} { \Gamma(\frac{\kappa+2}{2})} \left( \frac {\beta^{2-\kappa}(\kappa+1)}{g \kappa^2} \right)^{1/\kappa} \left[p^2-\delta+ \frac{\beta^2(\kappa-2)} {\kappa^2(\kappa+2) }\right] . 
\eq

For $\delta=0$, $\kappa>2$  these solitary waves are unstable  
\cite{stable1,cooper3}.  
For  $ \kappa=2$ the solitary wave is a critical one. In this case, 
the energy is {\it independent} of the width of the solitary wave.
Further, in the rest frame $v=0$, the energy is zero when $\delta=0$.  When $\delta$ is not zero there is a special constant  phase solution of  the form Eq. \eqref{exact}.  The condition for $\phi$ to be independent of $t$ is 
\bq
(p^2  +\beta^2/\kappa^2+\delta)=0\,,
\eq
which is only possible for negative $\delta$. 
For this solution, when $p=0$ we find the relationship:
\bq
\beta^2 = - \kappa^2  \delta. \label{constantphase1}
\eq
This solution will be the $r=0$ limit of some of the solutions  we will find below. 
\subsection{stability when $r=0$}
For the  unforced NLSE we can use the scaling argument of 
Derrick \cite{ref:derrick}  to determine if the solutions are unstable to scale transformation.  We have discussed this argument in 
our paper on the nonlinear Dirac equation \cite{NLDE}. 
The Hamiltonian  is given by 
\bq
H = \int dx  \left \{\frac{1}{2m} \psi^\star_x \psi_x - \delta \psi^\star \psi
-\frac{g^2}{\kappa+1} (\psi^\star \psi)^{\kappa+1} \right\} .
\eq
It is well known that using stability with respect to scale transformation to 
understand domains of stability applies to this type of Hamiltonian. 
This  Hamiltonian can be written as 
\bq 
H= H_1 - \delta  H_2- H_3 \>, 
\eq
where $H_i >0$.  Here $\delta$  can have either sign. 
If we  make a scale transformation on the solution which preserves the 
mass $M = \int  \psi^\star \psi dx$,
\bq
\psi_\beta \rightarrow \beta^{1/2} \psi(\beta x) \>,
\eq
we obtain
\bq
H=  \beta^2  H_1- \delta  H_2-  \beta^\kappa  H_3 \>.
\eq
The first derivative is: 
${\partial H}/{\partial \beta} = 2 \beta H_1 - \kappa  \beta^{ \kappa-1}  H_3$. 
Setting the derivative to zero at $\beta =1 $ gives an equation consistent 
with the equations of motion:
\bq
\kappa H_3 = 2 H_1 . 
\eq
The second derivative at $\beta =1$ can now be written as
\bq
\frac{\partial^2 H}{ \partial \beta^2} =   \kappa (2- \kappa) H_3  . \label{Derricktheorem}
\eq
The solution is therefore unstable to scale transformations when $\kappa >2$. 
This result is {\it independent} of $\delta$.  However once one adds forcing terms, it is known from the study of the $\kappa =1$ case
\cite{mqb} that the windows of stability as determined by the stability curve $p(v)$ as well as by simulation of the FNLSE increase as $\delta$  is chosen to be more negative. In those simulations the two methods agreed to within 1\%. 

\section{Exact Solutions of the Forced NLSE for $r  \neq 0$}

For $r=0$ one can have time dependent  phases in the traveling wave solutions, however when $r \neq 0$ one is restricted to looking for traveling wave solutions with time independent phases.    That is if we consider a solution of the form
\bq
 \psi(x,t) = e^{-i(kx+ \theta)} f(y);~~ y=x+2 kt ,
 \eq
  then $f$ satisfies
\bq
f^{\prime \prime}(y) - k'^2 f + g (f^\star f)^\kappa f = r ;  ~~k'^2 =  k^2 - \delta  .  \label{forced}
\eq

\subsection{Plane wave Solutions}
First let us consider the plane wave solution of Eq.( \ref {psieq}):
\bq
\psi(x,t) = a\exp{[-i(kx+\theta)]}\,,  \label{6}
\eq
or equivalently the constant solution $f=a$ to the Eq. (\ref{forced}).
This is a solution provided $a$ satisfies the equation: (here $a$ can be positive or negative)
\bq
r = g (a^{2 })^\kappa a  - (k^2-\delta) a . \label{r}
\eq
This solution has finite energy density, but not finite energy in general.  The energy density $H$ is given by
\bq
H=\frac{g(2 \kappa+1) a^{2(\kappa+1)}}{\kappa+1} - (k^2-\delta) a^2 . \label{plane}
\eq
Since $P_v =0$, the energy is the same in the comoving frame. 
This is the lowest energy solution for unrestricted $g$, $a$, $k^2$, and $\delta$  The solitary wave solutions we will discuss below will have finite energy with respect to this
``ground state" energy.  There is a special zero energy solution that is important. This solution has the restriction that:
\bq
\frac{g(2 \kappa+1) a^{2 \kappa}}{\kappa+1} =(k^2-\delta) . \label{zero}
\eq

\subsection{$2 \kappa= \text{integer}$}

When $2\kappa = {\rm integer}$ then one also can simply find solutions. The differential equation becomes
\bq
f'' -( k'^2)  f + g (f^*f )^\kappa f = r . 
\eq
Note that again  this equation is invariant under the combined transformation  $f \rightarrow -f$ and $r \rightarrow -r$. Thus we 
look for solutions of the form   $f_\pm =  \pm(a+u(y))$  with  $a>0; u(y) >0$ so that for this ansatz we have 
\bq
|a +u(y) | = a+ u(y) . 
\eq
It is sufficient to consider $f_+ = a+ u(y)$ and generate the second solution by symmetry ($ r \rightarrow - r; a \rightarrow -a; u \rightarrow-u $). For $f_+$
we have letting $2 \kappa +1 = N = \rm{integer}$
\bq
r = -k'^2 a + g a^N ,
\eq
and
\bq
u'' = (k'^2-gNa^{N-1} )u -  g\sum_{m=2}^N u^m a^{N-m}   \left(   \begin{array} {cc}
    N \\
    m\\ 
   \end{array}   \right).   
   \eq
  Integrating once and setting the integration constant to zero to keep the energy  of the solitary wave relative to a plane wave finite, we obtain
  \bq\label{N}
(u')^2 = (k'^2-gNa^{N-1} ) u^2  - g  \sum_{m=2}^N  \frac{ 2 u^{m+1} a^{N-m}   N!}{(m+1)! (N-m)!} .
\eq
This equation will have a solution as long as  $\alpha =k'^2-gNa^{N-1} > 0$.
We now discuss solutions to this equation when $N=2$, $ (\kappa=1/2)$ 
and $N=3 $, $(\kappa=1)$.

\subsection{$\kappa=1/2$}

For $\kappa=1/2$, $N=2$, the equation we need to solve for $f_+$  is
 \bq
(u')^2 =  \alpha  u^2  -  2 g a u^3 - g u^4/2; ~~ \alpha = k'^2 -2ga. 
\eq
This equation has a solution of the form 
\bq \label{f1/2}
u(y) = b ~\sech^{1/\kappa} (\beta y),
\eq
so that $f(y)$ has a solution of the form:
\bq \label{fy1}
f(y) =a+b\sech^2 \beta(y,t) , 
\eq
where both $ a, b >0,$ and $f$ obeys the equation:

\bq
f'' -( k'^2)  f + g |f|  f = r .  \label{joke}
\eq
The wave function is given by 
\bq \label{psi1}
\psi(x,t)= \exp[-i(kx+\theta)][a+b\sech^2 \beta(x+2kt)] . \,,
\eq

First consider the case that $a>0$ and $b>0$, so that $|f| =f$.  Substituting the ansatz Eq. (\ref{fy1}) into   Eq. (\ref{joke}) we obtain
\bq
a^2 g+2 a b g \text{sech}^2(\beta  y)-a \left(k'\right)^2+b^2 g \text{sech}^4(\beta 
   y)-b \left(k'\right)^2 \text{sech}^2(\beta  y)-6 b \beta ^2 \text{sech}^4(\beta 
   y)+4 b \beta ^2 \text{sech}^2(\beta  y)-r=0 .
   \eq
Equating powers of $sech$ we obtain the conditions:
\bq 
-k'^2 a+ ga^2 =r;~~ 4 \beta^2 -k'^2 + 2ga =0;~~ -6b \beta^2 +gb^2 =0 . \label{cond}
\eq
Or, equivalently 
\ba  
 a= \frac{k'^2 - \sqrt{k'^4+4rg} }{2g} , ~~~ 
 \beta^2 = \frac{1}{4} \sqrt{k'^4+4rg}, ~~~
 b = \frac{6 \beta^2}{g} . \label{re1}
\ea
The assumption  $a>0$ and $b>0$  requires for consistency   that $k'^2 >0$ and $r <0$.
So we can rewrite:
\ba  
 a= \frac{k'^2 - \sqrt{k'^4-4 |r| g} }{2g} , ~~~
 \beta^2 = \frac{1}{4} \sqrt{k'^4-4|r|g} , ~~~
&& b = \frac{6 \beta^2}{g}.\label{re2}
\ea
We also have the restriction that 
\bq
 \alpha = k'^2 - 2ga >0 . 
 \eq
The energy density corresponding to this solution is easily calculated and
is given by
\bq\label{11}
H=\frac{4ga^3}{3} -(k^2-\delta)a^2 +b^2[k^2-\delta-2ga+(2/3)bg]
\sech^4 [\beta(x+2kt)] -(4/3)gb^3 \sech^6 [\beta(x+2kt)] . 
\eq
Observe that the constant term is exactly the same as the energy 
density of the solution (\ref{6}) at $\kappa =1/2$. Hence the
energy of the pulse solution (over and above that of the solution 
(\ref{6})) is given by
\bq\label{12}
E = \frac{384 \beta^5}{5g^2}\,.
\eq

 Because of symmetry there is also a solution with $f \rightarrow -f , r \rightarrow  -r$. so that we can write the two  solutions connected by this symmetry  as follows:
\ba \label{fy3}
f(y)&& = - {\rm sign}(r) \left[ {a}+ {b}   \sech^2 (\beta y) , \right] \nonumber \\
\ea
with $a$, $\beta$ and $b$ given by Eq. (\ref{re2}).
Thus the wave function can be written as 
\bq
\psi(x,t) =  - {\rm sign}(r) \left[ {a}+ {b} \sech^2 \beta(x+2kt ) \right] e^{-i(kx+\theta)} . 
\eq
Notice that the boundary conditions (BC) on this solution are that for $\psi(x)$ at $\pm \infty$ the solution goes to the plane wave solution:
\bq
\psi(x=\pm \infty)  \rightarrow   - {\rm sign}(r) a \exp[-i(kx+\theta)] . 
\eq
Thus the BC on solving this equation numerically is the mixed boundary condition:
\bq \label{mixed}
ik \psi(x=\pm \infty,t) + \psi_x(x=\pm \infty),t) =0.
\eq
On the other hand if we go into the $y$ frame where 
\bq
u(y=\pm \infty,t)  \rightarrow     - {\rm sign}(r) a , 
\eq
then the BC for the $u$ equation is
\bq
u_{y} (y=\pm \infty,t)  = 0.
\eq

Consider the case where $g=2, k=-0.1, \delta = -1, r = -.01$. For this  choice of 
parameters  the exact  solution is given by
\bq
f(y)  = 0.727191 \sech^2(0.492338 y)+0.0101031 .   \label{exact1}
\eq

\begin{figure}[ht!]
\begin{center}
\includegraphics[width=7.0cm]{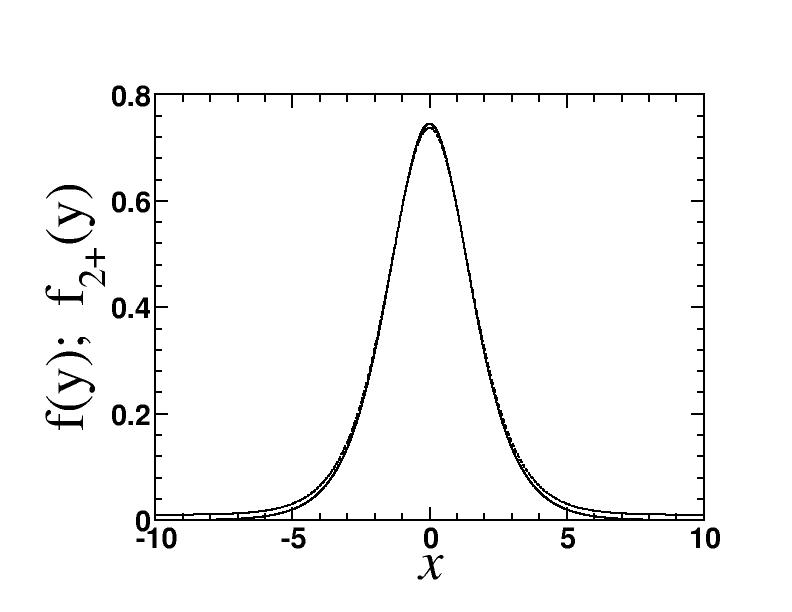} 
 \end{center}
\caption{Exact $f(y)$ (solid line) versus variational solution $f_{2+}(y)$ 
(dashed) for  $k=-0.1$, $\delta =-1$, $g=2$, $r=-0.01$.  
}
\label{newfig1} 
\end{figure}

The stationary solution found using  our variational ansatz which we will discuss below (See Eq.(\ref{varfplus}))  yields the approximate solution:
\bq
f_{2+}(y) =0.745042 \sech^2(0.498345 y) , 
\eq
which is a reasonable representation of the exact solution as seen in Fig.  \ref{newfig1}.
We will show later by numerical simulation, that  this solution is metastable in that $\beta$ remains constant for a short period of time and then oscillates.  In the numerical simulation of  Fig. \ref{s2c} the parameters used in the simulation are  $\kappa=1/2, \delta = -1, g=2, k=0.1, r=-0.075$.  For that case:
\bq
f(y) = 0.486113 \text{sech}^2(0.402539 y)+0.0904622 .   \label{reffigs2c}
\eq

\begin{figure}[ht!]
\begin{center}
\includegraphics[width=7.0cm]{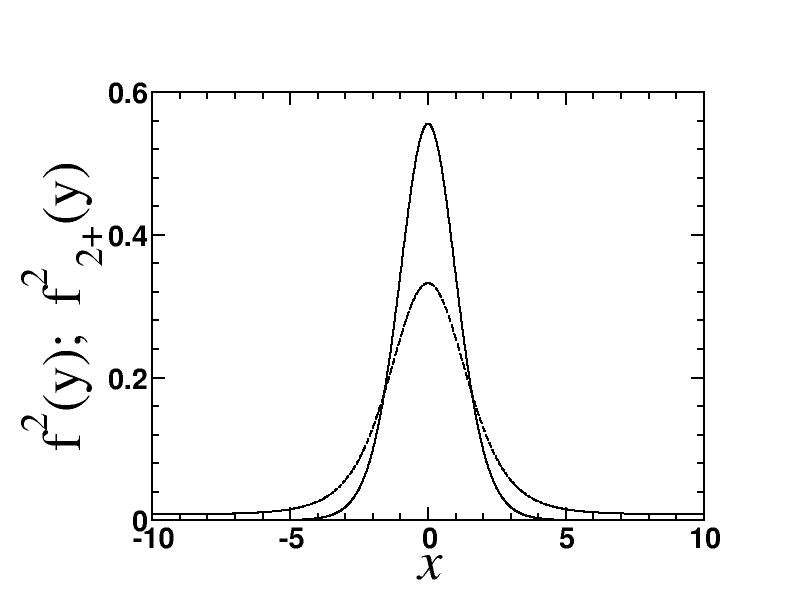} 
 \end{center}
\caption{Exact $f^{2}(y)$ (solid line) versus variational solution $f^2_{2+}(y)$  
(dashed line) for  $\delta = -1$, $g=2$, $k=0.1$, $r=-0.075$. 
These parameters were used in the simulations shown in Fig. \ref{s2c}.
}
\label{ugga} 
\end{figure}

Note that here $r$ is 14 \% of A so that the variational solution is worse.  The unstable stationary variational solution corresponding to this is 
\bq 
f_{2+}(y) =0.745535 \text{sech}^2(0.498509 y) .
\eq
To make a comparison we plot the square of these solutions in Fig. \ref{ugga}. 

\subsection{Finite energy solutions} 
\begin{figure}[ht!]
\begin{center}
\includegraphics[width=7.0cm]{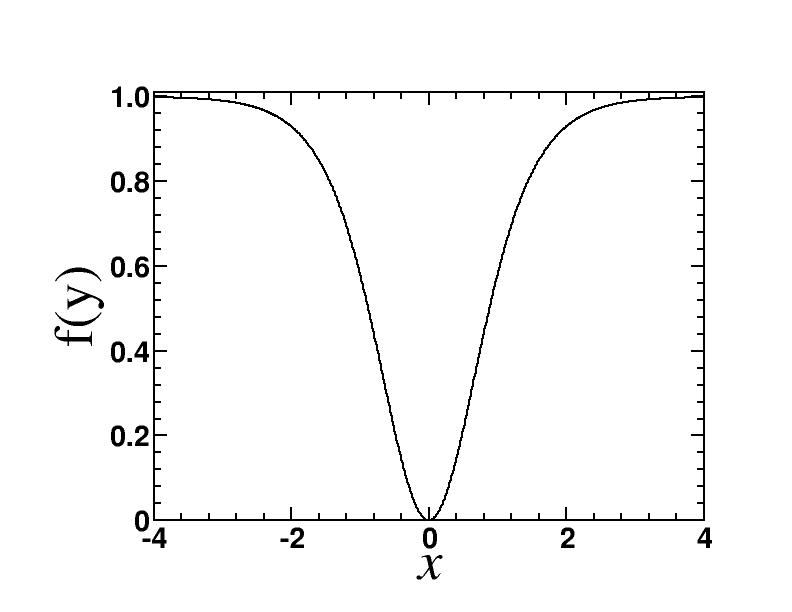} 
 \end{center}
\caption{Finite energy solution for $\kappa = 1/2$, $\beta =1$, 
normalized to unit height.
}
\label{figa} 
\end{figure}

~From Eq. \eqref{11} we notice that we can have  a finite energy solution 
(with energy as given by Eq. (\ref{12})) when the constant term in the 
energy density is zero.
This leads to the condition $4ga = 3  k'^2$, and yields $ b = -a$. 
This solution is {\it not} a small perturbation on the $r=0$ solution which has $a =0, b \neq 0$.  Instead for the finite energy solution
$a=-b$ and the form of the solution is 
\bq
f(y) = A \tanh^2 \beta y ,  \label{solf}
\eq
which has the appearance of Fig. \ref{figa}.

If we insert the solution Eq. (\ref{solf}) into Eq.  (\ref{joke}), we obtain

\bq
A g |A| \tanh ^4(\beta  y)+2 A \beta ^2-A k'^2 \tanh ^2(\beta  y)+6 A
   \beta ^2 \tanh ^4(\beta  y)-8 A \beta ^2 \tanh ^2(\beta  y)-r=0\,.
   \eq
   Thus Eq. (\ref{solf}) is a solution provided 
   
   \bq
   r= 2 A \beta^2;~~\beta^2 = - k'^2/8; ~~ |A|= - 6 \beta^2/g , 
   \eq 
   which requires that both  $g<0$ and $ k'^2<0$.  We see this  solution has the symmetry that $r$ changes sign with $A$ and thus  the sign of $f$ depends on the sign of $r$.   Because the coupling constant needs to be negative, the quantum version of this theory would not have a stable vacuum.    We can write this solution as 
   \bq
   f_ = \frac{r}{|r|}  \frac{6 \beta^2}{|g|}  \tanh^2 \beta y , 
   \eq
   where now
$\beta^2 = |k'^2|/8$ and  $r$ has the special values:
$r_\pm = \pm ({3}/{16}) {(k'^2)^2}/{|g|}$ . 
   
\subsubsection{Periodic Solutions for $\kappa=1/2$}

For $\kappa=1/2$ one  can easily verify that there are periodic solutions of Eq. (\ref{forced}) 
of  the form
\bq
f_+ (y) = b ~ {\rm dn}^2(\beta~ y,m) +a , 
\eq
where ${\rm dn}(x,m)$ is the Jacobi Elliptic Function (JEF)  with modulus $m$. 
Again for $a>0;, b>0$ we have $|f_+| = f_+$. 
Matching coefficients of powers of ${\rm dn}$ one obtains
\ba
&& a = \frac{\sqrt{1-m+m^2} k'^2 -(2-m) \sqrt{4 g r+k'^4}}{2 g
   \sqrt{1-m+m^2}}, \nonumber \\
 &&  b = \frac{6\beta^2}{g}\,,~~
 \beta^2 = \frac{\sqrt{k'^4+4 g r}}{4 \sqrt{1-m+m^2}}\, . 
\ea
The requirement  that $a>0$ translates into a restriction on $r$, namely $ r < -3(1-m)k'^4/4g(2-m)^2$.

\subsection{$\kappa =1$}
When $\kappa = 1$, the equation for $f$ is 
\bq
f'' - k'^2 f + g |f|^2  f =r . 
\eq
Again it is only necessary to consider $f_+ = ( a + u(y))$ with both $a>0$ and $u(y) >0$. 
For $\kappa=1$, $N=3$, the equation we need to solve for $u(y)$  is
 \bq
(u')^2 =  \alpha  u^2  -  2 g a u^3 - g u^4/2 . 
\eq
This equation has two solutions of the form
\bq
u(y) = \frac{b}{c  \pm \cosh(\beta y)},
\eq
 (note $y=x+2kt$).  These solutions were found earlier by Barashenkov and collaborators, \cite{bar1} \cite{bar2}.
Our condition on $u$ for $f_+$ requires that  $b>0, c>0$ and we choose the $+$  solution.  This leads to 
\bq
\label{15}
\psi_+ (x,t)= \exp[-i(kx+\theta)]\left[a+\frac{b}{c +  \cosh \beta(x+2kt)}\right]\,,
\eq
provided
\bq
\beta = \sqrt{\alpha};  ~~ b =  \frac{\sqrt{2}\alpha} {
\sqrt{2 g^2 a^2 + g \alpha}};~~ c = \frac{\sqrt{2} a g}
{\sqrt{2 a^2 g^2 + g \alpha}};~~  \alpha = k'^2 - 3 g a^2\,.
\eq
Since $\alpha >0$, we require $k'^2 > 3g a^2 >0 $.  Here $r=  -k'^2 a + g a^3  < - 2 g a^3 $is negative. 
When we let  $a \rightarrow -a, b\rightarrow -b, c  \rightarrow -c $,  then $r$ changes sign and becomes  positive.  Thus we can write 
\bq
\label{kappa1}
\psi_ \pm (x,t)=- \rm{sign (r)}  \exp[-i(kx+\theta)]\left[|a|+\frac{|b|}{|c| +  \cosh \beta(x+2kt)}\right]\, . 
\eq

These solutions are  non-singular since
\bq
1 - c^2 = g\alpha/[g\alpha+2g^2 a^2] > 0\,.
\eq

The energy density corresponding to these solutions is given by
\bq\label{17}
H=\frac{3ga^4}{2} -k'^2 a^2 
+\frac{2b^2(k'^2-3ga^2)}{[c+\cosh (\beta y)]^2}
-\frac{2b^2(c\beta^2+gab)}{[c+\cosh (\beta y)]^3}
+\frac{b^2[\beta^2(c^2-1)-(g/2)b^2]}{[c+\cosh (\beta y)]^4}\,.
\eq
Note that the constant term is exactly the same as the energy of the 
solution (\ref{6}) as given by Eq. (\ref{plane}) at $\kappa =1$. Hence the
energy of the $\kappa=1$ pulse solution (over and above that of the 
solution (\ref{6})) is again finite and given by
\bq\label{18}
E = \frac{8\alpha^{5/2}}{g\alpha+2a^2g^2}I_{2}-
\frac{16\sqrt{2}\alpha^{5/2}ag}{[g\alpha+2a^2g^2]^{3/2}}I_{3}-
\frac{8g\alpha^{7/2}}{[g\alpha+2a^2g^2]^2}I_{4}\,,
\eq
where
\bq\label{19}
I_{j} = \int_{0}^{\infty} \frac{dy}{[c+ \cosh (\beta y)]^{j}}\,,~~j=2,3,4.
\eq
We have 
\bq\label{20}
I_1=\int_{0}^{\infty} \frac{dx}{a+b\cosh(x)} =\frac{2}{\sqrt{b^2-a^2}}
\tan^{-1}\left[\frac{\sqrt{b^2-a^2}}{b+a}\right]\,,~~if~~ b^2 > a^2\,.
\eq
Note that in our case $a=c,b=1,c^2 < 1$. From here it is easy to calculate 
the integrals $I_{2},I_{3},I_{4}$ and hence show that
the energy of the $+$ and
the $-$ solution (over and above the solution (\ref{6})) is given by
\bq\label{24}
E = \frac{8\alpha^{1/2}}{3g}(\alpha+3a^2g) - \frac{8\sqrt{2}a}{\sqrt{g}}
(\alpha+2a^2g)
\tan^{-1}\left[\sqrt{1+\frac{2a^2g}{\alpha}}- \sqrt{\frac{2a^2g}{\alpha}}\right]\,.
\eq

\subsubsection{Periodic solutions for $\kappa =1$}

For $\kappa=1$ we can generalize the solitary wave solution for $f_+$  
\bq\label{k1}
f'' - k^2 f +g f^3 =r , 
\eq
namely
\bq
f= a + \frac{b}{c+\cosh[d y] }= \frac{(ac+b)  \sech[dy]+ a} { 1+ c ~\sech[dy]}
\eq
and obtain a periodic solution in terms of the Jacobi elliptic functions. The generalization of $\sech[dy]$ is the Jacobi elliptic function
$ \text{dn}(dy,m)$ which also has the property $| \text{dn}| =  \text{dn}(dy,m)$

One finds that 
\bq
f_+  = \frac{a + h ~ \text{dn}(dy,m)}{1+ c ~ \text{dn}(dy,m)}\,,
\eq
with $a>0;,h>0,c>0$ obeys $|f_+|= f_+$ and 
is an exact solution to Eq. (\ref{k1}) provided  
\bq
  r=\frac{(a c+h) \left(a g h-c k'^2\right)}{2 c^2}; ~~ 
d^2=-\frac{3 a g h-c k'^2}{2 c (m-2)} , 
  \eq
  \bq
h^2= \frac{4d^2}{g[2+\alpha g a^2(1-k'^2 \alpha)]}\,,~~
  (1-m)h^2  = \frac{1+ 2 \alpha g a^2-k'^2 \alpha}{4 d^2 \alpha^3 g^2 a^2}\,.
  \eq
Here $\alpha= {c}/{gah}$ which obeys a cubic equation:
  \bq
  16(1-m)\alpha^3 d^4 ga^2 = [2+\alpha g a^2(1-k'^2 \alpha)] 
[1+ 2 \alpha g a^2-k'^2 \alpha]\,.
  \eq
  
\subsection{$\kappa=3/2$}

For $\kappa = 3/2, N=4$ we have instead for $f_+$ that $u$ obeys the equation

 \bq
(u')^2 =  \alpha  u^2  -  4 g a^2 u^3 -  2  g a u^4 - 2  g u^5/5 . 
\eq

For this case,  the formal solution: 
\bq
 y + c = \int_a ^u dy \frac{1}{y \sqrt{ \alpha - 4 g a^2 y - 2 g a y^2 - 2 g  y^3/5}}
 \eq
 leads to an elliptic function. 

\section{Variational approach to the Forced NLSE}

For the problem of small perturbations to the unforced problem we are interested in variational trial wave functions of the form:
\bq
\psi_{v1}(x,t) = A(t) f[ \beta_v(t) \left(x-q(t) \right]  
e^{i \left[ p(t)\left(x-q(t) \right) + \phi(t)\right]}\,.  \label{varpsi}
\eq
Here we assume that the collective coordinates  (CCs)  $A(t),  \phi(t), p(t), q(t) $ are real functions of time and that $f$ is real. 
On substituting this trial function in Eq. (\ref{action}) and computing 
various integrals one finds that the effective Lagrangian is given by
\ba\label{efflag0}
&& L =  C_0 \frac{A^2(t)}{\beta_v}  \left(p(t) \dot{q}(t)-\dot {\phi}(t) -( p^2(t)-\delta)-  \frac{D_1}{C_0} \beta_v^2 \right) \nonumber \\
&& + C_k  \frac{g [A(t)]^{2(\kappa +1)} } {\beta_v(\kappa+1)} 
-4 r  A(t)  \cos[k q(t) + \phi(t)+ \theta] ~I[ p(t)+k, \beta_v]\,,
\ea
where
\bq
C_k = \int_{-\infty}^\infty~ dy [f(y)]^{2( \kappa +1)}
\,,~~ D_1 =\int_{-\infty}^\infty~  [f'(y)]^2\,, 
\eq
while
\bq\label{forcing}
I[p(t)+k],\beta_v]=  \int_0^{\infty} dy f(\beta_v y) \cos [(p(t)+k) y]\,. 
\eq

Note that for our parametrization of the variational ansatz: 
\bq
\psi^* \psi p(t) = \frac{1}{2i} \left( \psi^* \partial_x \psi - \psi \partial_x \psi^* \right). 
\eq
Thus from our previous discussion about the conservation of momentum in the comoving frame with $k=v/2$ we have that $p(t)$ and $M(t)$ are in general  not independent but satisfy 
\bq
M(t) ( p(t) + k) = {\rm constant} . 
\eq
For the case where $ p(t) + k \neq 0$, one then gets the relation
\bq
M(t)  = {\rm Constant} /( p(t) + k ) . 
\eq

\subsection{Variational Ansatz}
Generalizing the $\kappa=1$ choice of  \cite{Mpreprint} for $f$ to arbitrary $\kappa$, we choose
\bq
f_v[y] = \text{sech}^{1/\kappa} [y]  \label{choice1} , 
\eq
so that our variational ansatz is
\bq \label{ansatz}
\psi_{v1}(x,t) = A(t)  {\rm sech}^{1/\kappa}  [ \beta_v(t) \left(x-q(t) \right]  e^{i \left[ p(t)\left(x-q(t) \right) + \phi(t)\right]} , 
\eq
where $A(t)$ is of the same form as in the unforced solution, namely
\bq
A(t)= \left[ \frac{\beta_v(t)^2 (\kappa+1)}{g \kappa^2}\right]^{1/(2\kappa)} = \beta^{1/\kappa}\sqrt{ \alpha(\kappa)};~~  \alpha(\kappa) = \left(\frac{\kappa+1}{g \kappa^2} \right)^{1/\kappa}.
 \label{choicea}
\eq
Note that we have chosen $f$ to be real.  In our previous studies of blowup in the NLSE we used a  more complicated variational ansatz where there is another variational parameter in the phase multiplying the quadratic term $[x-q(t)]^2$ \cite{stable1}   \cite{cooper3}. 
Defining the ``mass"  of the solitary wave $M(t)$ via 
\bq 
M(t) = \int dx \psi^* \psi = C_0 A(t)^2 /\beta_v(t) = C_0 \beta^{2/\kappa -1} \alpha(\kappa),  \label{mass1}
\eq
we then find for  the ansatz Eq. (\ref{choicea})  that for $ p+k \neq 0$ 
\bq \label{rela}
  \beta^{\frac{2}{\kappa} - 1} (p(t) +k ) = \rm{constant} . 
  \eq

For the forcing term contribution to the Lagrangian, Eq. (5.2), we need  the integral
\begin{equation}
I[\nu,a,\beta_v]= \int_{0}^{\infty} dx \cos(ax) \sech^{\nu}(\beta_v x)
=[2^{\nu-2}/\beta_v\Gamma(\nu)] \Gamma (\nu /2+ia/2\beta_v)\Gamma(\nu/2-ia/2\beta_v)
\end{equation}
where $Re \beta_v > 0$, $Re(\nu) > 0$, $a >0$.   
Special cases of this are obtained for $\nu= 1,2,3$. 
We find
\bq
I[1,a, \beta_v] =\frac{\pi ~ \text{sech}\left(\frac{\pi a}{2 \beta_v}\right)}{2 \beta_v},~~
I[2,a, \beta_v]=\frac{\pi a ~\text{csch}\left(\frac{\pi  a}{2 \beta_v}\right)}{2 \beta_v^2} , 
~~I[3,a,\beta_v]=\frac{\pi  \left(a^2+\beta_v^2\right) \text{sech}\left(\frac{\pi  a}{2
   b}\right)}{4 \beta_v^3} . 
\eq

For our variational ansatz  Eq. \eqref{choice1}   
\bq  \label{params}
C_0 =  \frac{\sqrt{\pi } \Gamma \left(\frac{1}{\kappa }\right)}{\Gamma
   \left(\frac{1}{2}+\frac{1}{\kappa }\right)};~~
C_\kappa=\frac{\sqrt{\pi } \Gamma \left(1+\frac{1}{\kappa }\right)}{\Gamma
   \left(\frac{3}{2}+\frac{1}{\kappa }\right)}; ~~
D_1= \frac{\sqrt{\pi } \Gamma \left(\frac{1}{\kappa }\right)}{2 \kappa ^2
   \Gamma \left(\frac{3}{2}+\frac{1}{\kappa }\right)},
   \eq
 so that  
   \bq
   C_\kappa = \frac{2}{2 + \kappa} C_0;~~ D_1 =\frac{1}{\kappa(\kappa+2)}C_0\,.
   \eq
  Thus we can write everything in terms of $C_0$. 
 \subsection{Arbitrary $\kappa$}
   For arbitrary $\kappa$ the effective Lagrangian is 
   \ba
&& L =  M(t)  \left(p(t) \dot{q}(t)-\dot {\phi}(t)- (p^2(t)-\delta) + \beta_v ^2 \frac {(2-\kappa)} {\kappa^2 (2+ \kappa)}    \right) \nonumber \\
&&  -\frac{ 2^{\frac{1}{\kappa }}~r~  \beta_v^{1/\kappa -1}\sqrt{\alpha(\kappa)} \Gamma_{+}
    \Gamma_{-}\cos (k q(t)+\phi
   (t)+\theta )}{ \Gamma \left(\frac{1}{\kappa }\right)},  \label{efflag}
\ea
where 
\be
\Gamma_{\pm}=\Gamma \left(\frac{1}{2} \left(\frac{\pm i
   (k+p(t))}{\beta_v (t)}+\frac{1}{\kappa }\right)\right);~~ M(t) = C_0 \alpha(\kappa) \beta^{2/\kappa -1}. 
\eq
 Introducing the notation
 \bq
 C=\frac{\pi(k+p)}{2 \beta};~~ B = \phi + kq + \theta,
 \eq
an important  special  solution is obtained in the 
limit $C  \to 0$, i.e. $p(t)=-k$. In this case the Lagrangian becomes 
 \begin{eqnarray}
 L[C&=&0] =  M(t)  \left(-k \dot{q}(t)-\dot {\phi}(t)- (k^2-\delta) + \beta_v ^2 \frac {(2-\kappa)} {\kappa^2 (2+ \kappa)}    \right) \nonumber \\
&&  -\frac{ 2^{\frac{1}{\kappa }}~r~  \beta_v^{1/\kappa -1}\sqrt{\alpha(\kappa)} \Gamma^2(\frac{1}{2 \kappa})
\cos (B)}{ \Gamma \left(\frac{1}{\kappa }\right)}.  \label{czero}
\end{eqnarray}

This leads to the equations:
 \begin{eqnarray}
 \dot{q} &=& - 2k , \label{eqq7} \\
 \dot{\beta} &=& -\frac{ 2^{\frac{1}{\kappa }}~r~ \alpha_{\kappa}^{-1/2} 
\beta_{v}^{-1/\kappa+1} 
\Gamma(\frac{1}{2}+\frac{1}{\kappa}) \Gamma^{2}(\frac{1}{2\kappa})
    \sin (B)}{\sqrt{\pi} (\frac{2}{\kappa}-1) \Gamma^{2} 
\left(\frac{1}{\kappa }\right)}, \label{eqb7} \\
 \dot{\phi}&=& k^{2}+\delta+ \frac{\beta^2}{\kappa^2} -
\frac{ 2^{\frac{1}{\kappa }}~r~ \alpha_{\kappa}^{-1/2} 
\beta_{v}^{-1/\kappa} (1-\kappa)
\Gamma(\frac{1}{2}+\frac{1}{\kappa}) \Gamma^{2}(\frac{1}{2\kappa})
    \cos (B)}{(2-\kappa) \Gamma^{2} \left(\frac{1}{\kappa }\right)}. 
\label{eqphi7}
 \end{eqnarray}
 Assuming the ansatz $\beta=\beta_{s}$ and 
 $\phi(t)=\phi_s -\alpha_{s} t$ in Eqs. (\ref{eqb7})-(\ref{eqphi7}), where 
 $\beta_{s}$, $\alpha_{s}$ and $\phi_{s}$ are constant, we obtain  
 \bq
 \alpha_{s}=-2 k^2, \quad \phi_{s}=n \pi -k q_{0}-\theta,   
 \eq
 where $n$ is an integer and $\beta_{s}$ solution of 
 \bq
 k^{2}-\delta-  \frac{\beta_{s}^2}{\kappa^2} +(-1)^{n}
\frac{ 2^{\frac{1}{\kappa }}~r~ \alpha_{\kappa}^{-1/2} 
\beta_{s}^{-1/\kappa} (1-\kappa)
\Gamma(\frac{1}{2}+\frac{1}{\kappa}) \Gamma^{2}(\frac{1}{2\kappa})
    }{(2-\kappa) \Gamma^{2} \left(\frac{1}{\kappa }\right)}=0\,. 
 \eq

These equations become much simpler in the comoving frame, so we will now turn our attention to solving the problem in that frame.

\section{Variational Approach for the Action in the comoving frame}
In the comoving frame we  use the following  ansatz for the variational trial wave function:
\bq
v_{v1}(y,t) = A(t) f[ \beta_v(t) \left(y-\tq (t) \right]  e^{i \left[ \tp(t)\left(y-\tq(t) \right) + \tphi(t)\right]} . \label{varcom}
\eq
Comparing with our general ansatz we have the relations:
\bq
y= x+ 2 kt;  \tp = p+k, \tq = q+2kt , {\rm and}~ \tphi=  \phi+kq + \theta  .  \label{galileo}
\eq
On substituting these relations in the effective Lagrangian (\ref{efflag0}), 
the Lagrangian in the comoving frame takes the form 

\ba
&& L_2 =  C_0 \frac{A^2(t)}{\beta_v}  \left(\tp(t) \dot{\tq}(t)-\dot {\tphi}(t)
-( \tp^2(t)+k^2-\delta)+ \beta_v ^2  \frac{(2-\kappa)}{\kappa^2(2+\kappa)}\right) \nonumber \\
&& + C_k  \frac{g [A(t)]^{2(\kappa +1)} } {\beta_v(\kappa+1)} 
 -4 r  A(t)  \cos[\tphi(t)]  I[\tp(t),\beta]  . \label{efflag2}
\ea
We notice that the canonical momenta to $\tphi$ and $\tq$ are simply given by
\bq
\frac{\delta L_2}{\delta \dot{\tq}} = M(t) \tp(t);~ \frac{\delta L_2}{\delta \dot{\tphi}}= - M(t) . 
\eq
\subsection{Variational Ansatz function choice}
Again choosing  $ f_v[y] $  from Eq. \eqref{choice1}
our variational ansatz for $v$ is 
\bq \label{ta}
v_{v1}(y,t) = A(t) {\rm sech}^{1/\kappa}  [ \beta_v(t) 
\left(y-\tq(t) \right]  e^{i \left[ \tp(t)\left(y-\tq(t) \right) + \tphi(t)\right]} , 
\eq
where $A(t)$ is again given by Eq. \eqref{choicea}.  
Again  the ``mass"  of the solitary wave $M(t)$ is given by 
$M(t) = \int dx v^* v = C_0 A(t)^2 /\beta_v(t) = C_0 \beta^{\frac{2}{\kappa} - 1} \alpha(\kappa)$ and from the conservation of $P_v =M \tp$ (Eq. \eqref{pv}) 
we find for $\tp \neq 0$
\bq
  \beta^{\frac{2}{\kappa} - 1} \tp(t) = \rm{constant} . \label{rela1}
  \eq
We can rewrite Eq. \eqref{efflag2} as     
      \ba
&& L =  M(t)  \left(\tp(t) \dot{\tq}(t)-\dot {\tphi}(t)- (\tp^2(t)+k^2 -\delta) + \beta_v ^2 \frac {(2-\kappa)} {\kappa^2 (2+ \kappa)}    \right) \nonumber \\
&&  -\frac{ 2^{\frac{1}{\kappa }}~r~  \beta_v^{1/\kappa -1}\sqrt{\alpha(\kappa)} \Gamma_{+}
    \Gamma_{-}\cos (\tphi
   (t))}{ \Gamma \left(\frac{1}{\kappa }\right)},  \label{efflag3}
\ea
where 
$\Gamma_{\pm}=\Gamma \left(\frac{1}{2} \left(\frac{\pm i
   \tp(t)}{\beta_v (t)}+\frac{1}{\kappa }\right)\right)$. 
From Lagrange's equation for $\tq$ we obtain
\bq \label{rela3}
\frac{d}{dt} \left[M(t) \tp(t) \right] = 0,  ~\rightarrow
~\beta^{2/\kappa-1} \tp = \beta^{2/\kappa-1} [p(t)+k]=constant. 
\eq
Letting 
\bq
G(x=\tp/\beta) =\frac{ 2^{\frac{1}{\kappa }}~r~ \sqrt{\alpha(\kappa)} \Gamma_{+}
    \Gamma_{-}}      { \Gamma \left(\frac{1}{\kappa }\right)} , 
 \eq
 so that 
 \bq
 \frac{\partial G} {\partial \tp}  = \frac{G^\prime}{\beta},~~   \frac{\partial G} {\partial \beta }  = - \frac{ \tp G^\prime }{\beta^2} , 
 \eq
 we get the following differential equations:
 \bq
 {\dot \tq} = 2 \tp + \frac{1}{\beta M(t)}  G' \beta^{1/\kappa -1} \cos \tphi(t) , 
 \eq
 \bq
 M {\dot \beta} = G \beta^{1/\kappa} \frac{\kappa}{2-\kappa} \sin \tphi(t) , 
 \eq
 and for ${\dot \tphi}$ we have
 \ba
0= &&\frac{(2-\kappa)}{\kappa}  \left( \tp(t) \dot{\tq}(t)-\dot {\tphi}(t)- (\tp^2(t)+k^2 -\delta)   \right)  \nonumber \\
&& +   \beta_v ^2 \frac {(2-\kappa)} {\kappa^3 }+ \frac{\beta^{1/\kappa}}{M(t)} \left(\frac{\tp}{\beta^2} G^\prime - \frac{1-\kappa}{\kappa \beta} G \right) \cos \phi.  \label{dottphi1}
\ea

  \subsection{$\tp/\beta \rightarrow 0$}
  Introducing the notation
$ \tC={\pi \tp}/{2 \beta}$,  
a special stationary  solution is obtained in the 
limit $\tC  \to 0$, i.e. $\tp(t)=0$. In this case the Lagrangian (\ref{efflag3})
becomes 
 \begin{eqnarray}
 L[\tC&=&0] = 
\frac{ \sqrt{\pi} \Gamma \left(\frac{1}{\kappa }\right)
   \alpha_{\kappa} \beta_v (t)^{2/\kappa-1} } {\Gamma(\frac{1}{2} + \frac{1}{\kappa}) }  \left[  \delta -\dot{\tphi}(t) - k^2  + \beta_v ^2 \frac {(2-\kappa)} {\kappa^2 (2+ \kappa)}    \right] \nonumber \\
&&-\frac{ 2^{\frac{1}{\kappa }}~r~ \sqrt{\alpha_{\kappa}} 
\beta_{v}^{1/\kappa-1} 
\Gamma^{2}(\frac{1}{2\kappa})
    \cos (\tphi)}{\Gamma \left(\frac{1}{\kappa }\right)} .  
\end{eqnarray}

This leads to the equations:
 \begin{eqnarray}
 \dot{\tq} &=& 0, \label{eqq9} \\
 \dot{\beta} &=& -~r~A~ 
\frac{ \beta^{(\kappa-1)/\kappa}}{2-\kappa}
    \sin (\tphi), \label{eqb9} \\
 \dot{\tphi}&=& \delta-k^2+ \frac{\beta^2}{\kappa^2} - ~r~ B \beta^{-1/\kappa} \left( \frac{1-\kappa}{2-\kappa}  \right)    \cos (\tphi) , 
\label{eqphi9}
 \end{eqnarray}
 
 where 
 \ba
 A&&= \frac{ 2^{\frac{1}{\kappa }}~ \kappa~  \alpha_{\kappa}^{-1/2} 
\Gamma(\frac{1}{2}+\frac{1}{\kappa}) \Gamma^{2}(\frac{1}{2\kappa})
  }{\sqrt{\pi}  \Gamma^{2} 
\left(\frac{1}{\kappa }\right)}, ~~A>0, \nonumber \\
 B&& =
\frac{ 2^{\frac{1}{\kappa }}~ \alpha_{\kappa}^{-1/2} 
\Gamma(\frac{1}{2}+\frac{1}{\kappa}) \Gamma^{2}(\frac{1}{2\kappa})}{ \sqrt{\pi}  \Gamma^{2} \left(\frac{1}{\kappa }\right)}; ~~B>0.
 \ea
 Assuming the ansatz $\beta=\beta_{s}$ and 
 $\tphi(t)=\tphi_s$ in Eqs. (\ref{eqb9})-(\ref{eqphi9}), where 
 $\beta_{s}$, and $\tphi_{s}$ are constant, we obtain  
 \bq
 \tphi_{s}=n \pi  , 
 \eq
 where $n$ is an integer and $\beta_{s}$ the solution of 
 \bq
 k^{2}-\delta-  \frac{\beta_{s}^2}{\kappa^2} +(-1)^{n}
~r~ B~ 
\beta^{-1/\kappa}~  \frac{ (1-\kappa)}{(2-\kappa)}
=0. 
 \eq

\subsubsection{Linear Stability}

Let us look at the problem of keeping $\tq=\tp =0$, and looking at small 
perturbations about  the stationary solutions
$\beta= \beta_s$ and $\tphi = \tphi_s =  0, \pi$.  
The analysis depends on whether $r \cos \tphi_s >0$ or $r \cos \tphi_s <0$. 
We discuss the two cases separately. We let $\rho =| r \cos \tphi_s | >0$. 

{\bf Case IA: $ r \cos \tphi_s  =  \rho > 0$}

The linearized equations for $\delta \beta, \delta \phi$ then become
\bq
\delta{\dot \beta} = - \frac{A~ \rho  \beta_s^{(\kappa-1)/\kappa}}{2-\kappa} 
\delta \tphi  = -c_1 \delta \tphi\,,
\eq
\bq
\delta{\dot \tphi } = \left[ \frac{2 \beta_s}{\kappa^2}+ \frac{ \rho ~B(1-\kappa)}
{\kappa(2-\kappa)\beta_s^{(\kappa+1)/\kappa}}\right] \delta \beta
= c_2 \delta \beta\,.
\eq
Combining these two equations by taking one more derivative we find:
\bq
 \delta {\ddot  \beta}+ \Omega^2 \delta \beta =  \delta {\ddot  \tphi}+ \Omega^2 \delta \tphi =0,
\eq
where $\Omega^2 = c_1 c_2$. Whether this will correspond to a stable solution 
will depend on signs of $c_1,c_2$ and hence $c_1 c_2$. It turns out that the
answer depends on the value of $\kappa$. In particular, the answer depends on
whether $\kappa \le 1$ or $ 1 < \kappa <2$ or $ \kappa >2$. So let us discuss 
all three cases one by one. Note that the above analysis is only valid if 
$\kappa \ne 2$. If $\kappa =2$, the entire analysis needs to be redone.

{\bf $\kappa \le 1$}: In this case $c_1, c_2 > 0$ and hence $\Omega^2 >0$
so that the solution is a stable one.

{\bf $1 < \kappa < 2$}: In this case, while $c_1>0$, the sign of $c_2$ will
depend on the values of the parameters. In particular, if $\kappa$ is 
sufficiently close to (but greater than) one, then $c_2 >0$ and 
hence $\Omega^2 > 0$ so that
one has a stable solution. On the other hand if $\beta$ is sufficiently 
close to (but less than) two, then $c_2 < 0$ and hence $\Omega^2 < 0$ so 
that one has an unstable solution. In particular, no matter what the values of 
the parameters are, as $\kappa$ increases from one to two, the solution will
change from a stable to an unstable solution. 

{\bf $2 < \kappa$}: In this case, while $c_1 < 0$, $c_2 > 0$ and hence
$\Omega^2 < 0$ so that one has an unstable solution. 

{\bf Case IB: $ r \cos \phi_s =-  \rho  <0. $}

The linearized equations for $\delta \beta, \delta \tphi$ then become
\bq
\delta{\dot \beta} = + \frac{\rho ~ A \beta_s^{(\kappa-1)/\kappa}}{2-\kappa} 
\delta \tphi  = c_1 \delta \tphi\,,
\eq
\bq
\delta{\dot \tphi } = \left[ \frac{2 \beta_s}{\kappa^2}- \frac{\rho ~B(1-\kappa)}
{\kappa(2-\kappa)\beta_s^{(\kappa+1)/\kappa}}\right] \delta \beta
= c_3 \delta \beta\,.
\eq
Combining these two equations by taking one more derivative we find:
\bq
 \delta {\ddot  \beta}- \Omega^2 \delta \beta =  \delta {\ddot  \tphi}- \Omega^2 \delta \tphi =0,
\eq
where $\Omega^2 = c_1 c_3$. Whether this will correspond to a stable solution 
will depend on signs of $c_1,c_3$ and hence $c_1 c_3$. 
Again it turns out that the answer depends on
whether $\kappa \le 1$ or $ 1 < \kappa <2$ or $ \kappa >2$. So let us discuss 
all three cases one by one. 

{\bf $\kappa \le 1$}: In this case $c_1>0$ while the sign of $c_3$ and 
hence $\Omega^2$ depends on the values of the parameters. For example if
$\kappa$ is sufficiently close to one, then $c_3 >0$ 
so that the solution is an unstable one.

{\bf $1 < \kappa < 2$}: In this case, both $c_1, c_3>0$, and hence 
$\Omega^2 >0$ so that the solution is an unstable one. 

{\bf $2 < \kappa$}: In this case, while $c_1 < 0$, the value of $c_3$
will depend on the value of the parameters. In particular if $\kappa$ 
is sufficiently close to two then $c_3< 0$ and hence
$\Omega^2 > 0$ so that one has a stable solution. On the other hand for
$\kappa > \kappa_c > 2$, $c_3 >0$ and 
hence $\Omega^2 < 0$ so that one has an unstable solution.  
In particular, no matter what the values of 
the parameters are, as $\kappa$ increases from two, the solution will
change from a stable to an unstable solution. 

We now discuss the $\kappa=1/2$ case in some detail.
   
   \subsection{$\kappa=1/2$}
   For $\kappa=1/2$ the effective Lagrangian is 
   \bq\label{k12}
   \frac{12}{g^2}  \left(- \pi  g r \tp(t) \text{csch}\left(\frac{\pi 
   \tp(t)}{2 \beta (t)}\right) \cos (\tphi (t) )+4 \beta
   (t)^3 \left(\tp(t) \dot{\tq}(t)-\tp(t)^2- \dot{\tphi} (t)+\delta-k^{2} 
\right)+
   \frac{48}{5} \beta
   (t)^5\right) . 
   \eq 
From the Euler-Lagrange equations we find that:
   \bq
\beta^3(t)\tp(t) = constant\,,   \label{constraint1}
\eq
and further,
\bq \label{qdot} \dot{\tq} = 2\tp + \frac{g \pi r \cos \tphi}{4 \beta^3 \sinh C} 
\left[1- C \coth C \right]  ;  ~~C = \frac{\pi \tp(t)}{2 \beta(t)} , 
\eq
\bq \label{betadot}
\dot{\beta} = - \frac{ gr C \sin \tphi}{6 \beta \sinh C} , 
\eq
\bq \label{pdot}
\dot{\tp} = \frac{ g r C^2 \sin \tphi}{\pi \beta \sinh C} , 
\eq
\bq \label{phidot}
\dot{\tphi} = \tp^2 +4 \beta^2 + \delta-k^{2} +  
\frac{g \pi r \tp  \cos \tphi}{4 \beta^3 \sinh C} \left[1-C \coth C \right] -\frac{gr C^2 \coth C \cos \tphi}{6 \beta^2 \sinh C}.
\eq
We are interested in the stationary solutions of these equations.  We assume
 \bq
 \tilde{q}= v_s t, ~~\beta = \beta_s, ~~\tp=\tp_s, ~~\tphi  = \tphi_s . 
 \eq
 We now have $C_s = \frac{\pi}{2 \beta_s} (\tp_s)$ and we need $\sin \tphi_s  =0$ so that  $ \tphi_s= 0,\pi$ and $\cos \tphi_s = \pm 1$.
 
 The equations for the two choices of $\cos \tphi_s $  become:
\bq \label{q1c } v_s = 2p_s \pm \frac {g \pi r }{4 \beta_s^3 \sinh C_s} \left[1- C_s \coth C_s \right] ,
\eq
 \bq \label{phi1c}
0   = \tp_s^2 +4 \beta_s ^2 - k'^2   \pm  \frac{g \pi r \tp_s  }{4 \beta_s^3 \sinh C_s} \left[1-C_s \coth C_s \right]  \mp \frac{gr C_s^2 \coth C_s}{6 \beta_s^2 \sinh C_s} . 
\eq
This leads to two families of stationary solutions on two curves in the three dimensional parameter space of  $v_s, \beta_s,$ and $ \tp_s$.
Equation \eqref{phi1c}  can also be written:
\bq
 \tp_s^2 = v_s p_s +  4 \beta_s ^2 - k'^2   \mp \frac{gr C_s^2 \coth C_s}{6 \beta_s^2 \sinh C_s} . 
\eq

\subsubsection{phase portrait}

\begin{figure}[ht!]
\begin{center}
\includegraphics[width=14.0cm]{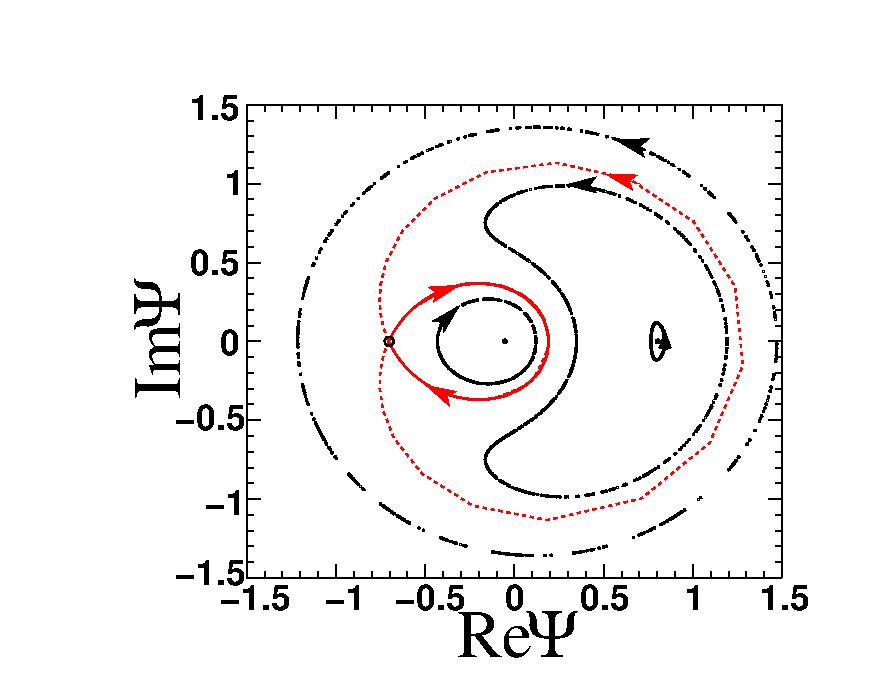}   
\end{center}
\caption{Phase portrait: imaginary vs. real part of the wave function Eq.\ 
(\ref{pp}). Orbits with positive sense of rotation predict stable solitons 
in the simulations. 
If the orbit, or part of it, has a negative sense, the soliton is unstable. 
The filled and open circles are stable and unstable fixed points, respectively, 
Eq.\ (\ref{fp}). The orbits are obtained for 
$\beta_{0}=0.2$ (medium size ellipse), 
$\beta_{0}=0.53$ (small ellipse), 
$\beta_{0}=0.63$ (horseshoe) and 
$\beta_{0}=0.7$ (large ellipse), keeping fixed $\phi_0=0$, $p_0=-k+10^{-5}$,  
$q_0=0$, except for $\phi_{0}=\pi$ 
and $\beta_{0}=0.498345$ (separatrix, red curve), see Fig.\ \ref{fig19A}. 
The parameters are $r=0.05$, $k=-0.1$, $\delta=-1$, 
$\theta_{0}=0$ and $\alpha=0$.}
\label{phaseportraita} 
\end{figure}

The variational wave function for $\kappa=1/2$ in the comoving frame  is given by 
\bq
v_v(y,t) = \frac{6 \beta^2}{g}  \sech^2 \beta \left[y-\tq(t)\right]  e^{i \left[\tp(t) (y-\tq(t) + \tphi(t) \right]} .
\eq
The position of the solitary wave $\tq(t)$ consists of a linear term $\bar{v} t$ 
plus an oscillating term. 
The average velocity $\bar{v}$ can be obtained from the 
numerical solution of the ODEs (\ref{qdot})-(\ref{phidot}) for the collective variables. 
The relevant wave function for the phase portrait \cite{Mpreprint}  
is to evaluate $v_v$ at the point $y= \bar{v} t$ . The phase portrait is obtained 
by  plotting  the real versus imaginary part of the resulting wave function as a 
function of time by solving the ODEs for various initial values of $\beta$.   
That is, we consider
\bq \label{pp}
\Psi= \frac{6 \beta^2}{g}  \sech^2 \beta(t) \left[\bar{v}t -\tq(t)\right]  e^{i \left[\tp(t) \left( \bar{v}t -\tq(t) \right)+ \tphi(t) \right]} , 
\eq
and plot  Im $\Psi$ vs.  Re $\Psi$  for fixed parameters varying the value of 
$\beta_0$. Orbits which 
are ellipses in the positive (negative)  sense of rotation predict stable 
(unstable)  solitary waves in the simulations.  When the orbit has both senses of 
rotation, as in a horseshoe shape, the solitary wave is unstable.  These behaviors are shown in the phase portrait of Fig. 
\ref{phaseportraita}. 
The three stationary solutions that we found earlier correspond to the fixed 
points of the phase portrait. The stability of these fixed points can be 
determined either numerically by solving the CC equations, or
analytically by a linear stability analysis that we will now discuss. 
All the fixed points are on the real axis
since $\tilde{\phi}_s = 0, \pi$ for the stationary solutions. 
We have at the fixed points
\bq \label{fp}
\Psi_s = \frac{6 \beta_s^2}{g}  e^{i  \tphi_s }.
\eq
We remark that a stable fixed point only means that the CC solutions 
in its neighborhood are stable, but an orbit around this fixed point 
does not necessarily have a positive sense of rotation. 
An example is the medium size ellipse in Fig.\  \ref{phaseportraita}.

\subsubsection{Linear stability analysis of stationary solutions}
To study the stability of these stationary solutions we let
\bq
\beta(t) = \beta_s + \delta \beta(t) ; ~~ \tphi(t) =  \tphi_s  + \delta \tphi(t) ; ~~\tphi_s = 0, \pi .
\eq
From Eq. \eqref{betadot} we obtain:
\bq
\delta {\dot \beta} = \mp  \frac{gr C_s^2}{6  \beta_s \sinh C_s} \delta \tphi = \mp c_\beta \delta \tphi  . \label{deltabeta}
\eq

We can schematically write Eq. \eqref{phidot} as follows:
\bq \label{phidota }
\dot{\tphi} = \tp^2 +4 \beta^2 -k'^2  +  F[ p, \beta, C] \cos \tphi , 
\eq
where
\bq
F[ p, \beta, C] = \frac{g \pi r \tp }{4 \beta^3 \sinh C} \left[1-C \coth C \right] -\frac{gr C^2 \coth C }{6 \beta^2 \sinh C}.
\eq

This leads to the equation for linear stability
\bq
\delta {\dot \tphi}  = 2 p_s \delta \tp  + 8 \beta_s \delta \beta \pm \left( \frac{\partial F}{\partial \tp} \delta \tp + \frac{\partial F}{\partial \beta } \delta \beta +\frac{\partial F}{\partial C } \delta C \right) ,  
\eq
where the derivatives are evaluated at the stationary values $\beta_s, C_s, \tp_s$
Now the conservation of momentum in the comoving frame leads to 
\bq
\beta^3 \tp = C_1 = constant. 
\eq
So we have
\bq
\tp =\frac{C_1}{\beta^3} \rightarrow \delta \tp = - 3  \frac{C_1} {\beta_s^4} \delta \beta , 
\eq
\bq
C=  \frac{\pi C_1}{2 \beta^4} \rightarrow \delta C = - 4  \frac{C_s} {\beta_s} \delta \beta . 
\eq

Putting this together we get:
\bq
\delta {\dot \tphi} = c_\phi \delta \beta   \label{deltaphi} , 
\eq
where
\bq
c_\phi =  - \frac{3 C_1}{\beta_s^4} (2 \tp_s  \pm   \frac{\partial F}{\partial \tp}) + 8 \beta_s \pm  \left(\frac{\partial F}{\partial \beta }  - \frac{4 C_s}{\beta_s} \frac{\partial F}{\partial C } \right) . 
\eq
We can write
\bq 
\delta {\dot \beta} = \mp c_\beta \delta \tphi , \label{deltabeta2} 
\eq
so taking derivatives of   Eqs \eqref{deltaphi} and \eqref {deltabeta2} and combining we get
\bq
 \delta \ddot{ \tphi}  \pm c_\beta c_\phi \delta \tphi  =0 , ~~~
 \delta \ddot{ \beta } \pm  c_\beta c_\phi \delta \beta =0 .
\eq
Thus, depending on the sign of $c_\beta  c_\phi$ we will have oscillating or growing (decreasing) solutions.

Once we have solved for $\delta \beta$, we can go back and solve for $\delta q$.  We can write the Eq. \eqref{qdot}  for $\dot{\tq}$ as
\bq \label{qdota} \dot{\tq} = 2\tp + \cos \tphi ~ F_1(\beta, C) , 
\eq
where 
\bq
 F_1 =  \frac{g \pi r }{4 \beta^3 \sinh C} . 
\left[1- C \coth C \right]
\eq
Letting $ \dot{\tq} = v_s + \delta \dot{\tq}$ we have
\ba
\delta \dot{\tq} && = 2 \delta \tp \pm   \left(\frac{\partial F}{\partial \beta }  \delta \beta  +\frac{\partial F}{\partial C } \delta C  \right) \nonumber \\
&& = \left[ -6 \frac{C_s}{\beta_s^4} \pm  \left(\frac{\partial F_1}{\partial \beta }  - \frac{4 C_s}{\beta_s} \frac{\partial F_1}{\partial C } \right) \right]  \delta \beta . \nonumber \\
&& = c_{q\pm} \delta \beta
\ea
Thus if we are in a region of stability so that 
\bq
\delta \beta =\varepsilon_\beta \cos (\Omega t + \alpha) , 
\eq
then we obtain
\bq
\delta \tq = \delta\tq(0) +  \frac{c_{q \pm} \varepsilon_\beta}{\Omega} \left[ \sin( \Omega t + \alpha) - \sin \alpha \right].
\eq

\subsection{Dynamical stability using the stability curve $p(v)$}
In references \cite{mqb} \cite{Mpreprint} it was shown for the case 
$\kappa =1 $ that the stability of the solitary wave could be inferred 
from the solution of the CC equations by studying the stability curve 
$p(v)$, obtained from the parametric representation 
$p(t), v(t) = \dot{q}$, where $p$ is the normalized canonical momentum in the lab frame and
$v= \dot{q}$ is the velocity in the lab frame. We can determine these quantities using the 
relations
\bq
p(t) = \tp -k; ~~  \dot{q} = \dot{\tq} - 2k .
\eq
A positive slope of $p(v)$ curve is a necessary condition for the stability of the solitary wave. 
If a branch of the $p(v)$ curve has a negative slope, this 
is a sufficient condition for instability.   In our simulations we will show that
this criterion agrees with the phase portrait analysis.

\subsection{ $\kappa = \frac{1}{2}, C=0$}
For the special case that $C=0$, using the identity 
$x~\text{csch} x \to 1$, the effective Lagrangian Eq. (\ref{k12}) becomes:
\bq
L[C=0] =-2 g r \beta (t) \cos (\tphi (t))+4 \beta (t)^3
   \left(\delta-k^{2} -{\dot \tphi} 
(t)\right)+\frac{48 \beta
   (t)^5}{5} . 
   \eq
This leads to the equations:
\begin{eqnarray}
\tp&=&0   ;  \dot{\tq} = 0,\nonumber \\
\dot{\beta} &=& - \frac{g r}{6 \beta} \sin (\tphi); ~~ \dot{\tphi} = \delta-k^2+ 4 \beta^2 - \frac{g r}{6 \beta^2} \cos(\tphi). \label{tp0} \\ 
\end{eqnarray}

The stationary solution in the comoving frame is represented 
by  $\beta=\beta_{s}$ , $\tq = q_0 $ and 
$\tphi(t)=\tphi_s$, where $\tphi_{s}$ is  given by  
\bq
\tphi_{s}=n \pi,   
\eq
with $n$ being an integer. It is sufficient if we take $n=0,1$. 
For $n=0$  and $r>0$ or equivalently, $n=1, r <0$ we have 
\bq
\beta_{s}^{2}=\frac{k'^{2}}{8} + 
\frac{\sqrt{k'^{4}+8 g|r| /3}}{8}, 
\eq
whereas for $r<0, n=0$  or equivalently $r>0, n=1$ there are two solutions given by  
\bq
\beta_{s}^{2}=\frac{k'^{2}}{8} \pm  
\frac{\sqrt{k'^{4}-8 g|r|/3}}{8}, \quad k'^{4}/(g|r|) > 8/3\approx 2.66667.  \label{posneg}
\eq
Note when $C=0$ then $p_s = 0$. 
We expect that these stationary solutions are close to exact solitary wave solutions to the original partial differential equations that may be
stable or unstable. 

We are interested in seeing how these solutions compare to the exact solution 
we found earlier as well as to the unperturbed $r=0$ exact solution.  The unperturbed solution with $r=0, p=0$ is given by
\bq
f_0(y) = 6  \frac{\beta^2}{g} \sech^2 [\beta y] , 
\eq
where $\beta^2 = - \delta/4$.  Thus for $\delta = -1, g=2$ the exact unperturbed solution is 
\bq
f_0(y) =\frac{3}{4} \text{sech}^2\left(\frac{y}{2}\right)  . \label{fzero}
\eq

The exact perturbed  solution  for $ r= -.01$ and $k=-0.1, \delta = -1$, is given by Eq.(\ref{exact1}) i.e.
$
f(y) = 0.727191 \text{sech}^2(0.492338 y)+0.010103.
$

Let us now look at the three stationary variational solutions. Since $r <0$ we have  for $n=0$ the two possibilities based on the choice of the $\pm$  in the square root.  For the positive choice we obtain:
\bq
f_+(y) =0.745535 \text{sech}^2(0.498345 y)  , \label{varfplus}
\eq
which is very close to $f_0$.  We will find from our linear stability analysis that this is unstable.

On the other hand, for the negative root we obtain
\bq
f_{-}(y)= 0.011965 \text{sech}^2(0.0631532 y) , 
\eq
which is far from $f_0$.  This solution is stable to small perturbations. 
For $n=1$ we obtain instead
\bq
\beta_{s}^{2}=\frac{k'^{2}}{8} +  
\frac{\sqrt{k'^{4}+8 g|r|/3}}{8}=  0.255148 , 
\eq 
so that
\bq
f_1(y) = 0.765443 \text{sech}^2(0.505121 y).  \label{mugga1}
\eq
We find that this is stable to small perturbations.

\begin{figure}[ht!]
\begin{center}
\includegraphics[width=7.0cm] {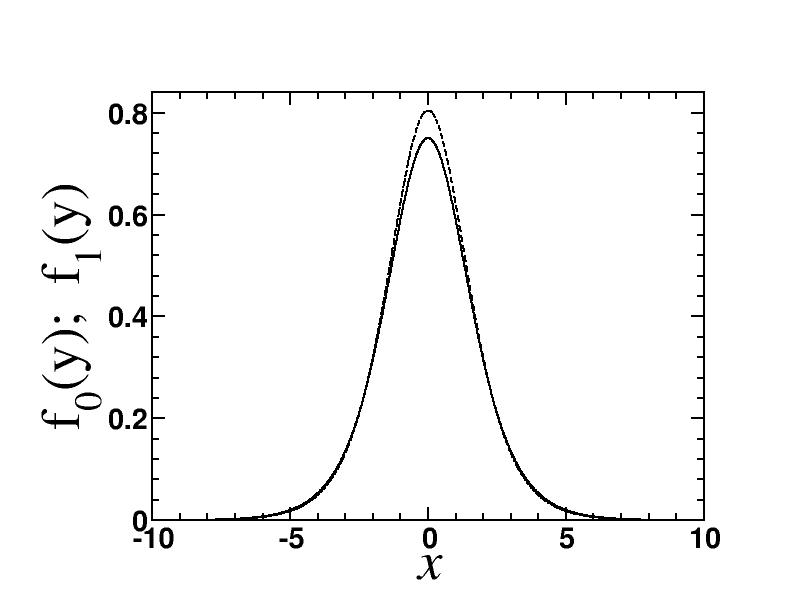} 
 \end{center}
\caption{$f_0(y)$  vs. forced variational solution for  $f_1(y)$  for  
$k=-0.1$, $\delta =-1$, $g=2$, $r=0.05$.}
\label{f1y} 
\end{figure}

 We are also interested for comparison purposes in using the same parameters we  used in the study of 
the  $\kappa=1$ problem \cite{Mpreprint}, i.e. $k=-.1, \delta= -1, g=2, 
r=0.05, q_0=0, \phi_s = 0$. Then for this positive value of $r$, the  
amplitude of the  $n=0$ solution is (note $y=x+2kt$) 
\bq
 f_1(y) = 0.804134 \sech^2 [0.51773 y]\,.  \label{f1c}
 \eq
 The phase of the solution is zero.  This is not very far from the value of 
$f_0$ given in Eq. (\ref{fzero}).
 The comparison is shown in Fig. \ref{f1y}. This solution turns out to be stable under small perturbations. If we instead chose $k = +0.1$,
 the result for $f_1(y)$ would be the same but the solitary wave would move in the opposite direction. 

The two solutions for $n=1, 
\tilde{\phi}_{s}= \pi$  are
\bq
 f_{2+}(y) = 0.704252 \sech^2(0.484511 y) ; ~~ f_{2-}(y) = 0.053248 \sech^2(0.133227 y) .  \label{f2}
\eq
$f_{2+}(y)$ is again similar to $f_0$ but   $f_{2-}(y)$ is clearly not.  Our linear stability analysis leads to the conclusion that 
 $f_{2+}(y)$ is unstable but  $f_{2-}(y)$ is stable.  In fact, $f_{2+} (y)$ is unstable to linear perturbations but is actually metastable and the original solution switches to a separatrix solution at late times.  $f_{2-}$ is stable to small perturbations. 
The separatrix for these values of the parameters is shown below  in  Fig. \ref{fig19A}.

Next we want to consider initial conditions when $r=0.005$ and $k= -0.01, \delta = -1, g=2$. For these initial conditions it is easier to deal with the periodic boundary conditions on the numerical solutions of the PDEs.  
The  $n=0$ solution is 
\bq 
f_1(y) =0.755042 \sech^2(0.501678 y)\,,
\eq 
and we find that this is stable to linear perturbations. 
The positive and negative roots for the $n=1$ solution are
\bq  f_{2+}(y) = 0.745042 \sech^2(0.498345 y) ; ~~f_{2-} (y) = 0.00503328 \sech^2(0.0409604 y) . \label{f2005}
\eq
We will find using linear stability analysis that  $f_{2+}(y)$ is unstable but  $f_{2-}(y)$ is stable. 

\subsection{linear stability at $\kappa = \frac{1}{2}, C=0$}

Let us look at the problem of keeping $\dot{\tq}=\tp =0$, and consider at small 
perturbation of Eqs. (\ref{tp0})  about  the stationary solution
$\beta= \beta_s$ and $\tilde{\phi} = \tilde{\phi}_s =  0, \pi$. 
 When $\tilde{\phi}_s =0$, 
the linearized equations for $\delta \beta, \delta \phi$ become
\bq
\delta{\dot \beta} = - \frac{gr}{6 \beta_s} \delta \tilde{\phi}  = 
-c_1 \delta \tilde{\phi} .  
\eq
Here the sign of $c_1$ is given by the sign of $r$. 
\bq
\delta{\dot \tilde{\phi} } = \left( 8 \beta_s+ \frac{gr} {3 \beta_s^3} \right) \delta \beta= c_2 \delta \beta ~~ c_2 > 0 . 
\eq
Combining these two equations by taking one more derivative we find:
\bq
 \delta {\ddot  \beta}+ \Omega^2 \delta \beta =  \delta {\ddot  \tilde{\phi}}+ 
\Omega^2 \delta \tilde{\phi} =0,
\eq
where $\Omega^2 = c_1 c_2 $.
Thus, for the case when we have an exact solution and we choose  $ r = -0.01, g=2, k=-0.1, \delta =-1$ we find that  for the stationary solution $f_+(y)$ we obtain
$c_1=-0.0066866$, $c_2=3.93426$, and $\Omega^2 = -0.0263068$, 
suggesting an unstable solution.
For $f_-(y)$ we instead obtain
$c_1=-0.0527817$, $c_2=-2.56198$, and $\Omega^2 =0.135226$, 
suggesting a stable solution.  When $r>0$, the solution  $f_1(y)$  
which corresponds to $r>0, \phi = 0$ is always stable.

If instead we look at the small perturbations around the solution where $\tilde{\phi}_s = \pi$, we instead obtain
\bq
\delta{\dot \beta} = \frac{gr}{6 \beta_s} \delta \tilde{\phi}  = c_1 \delta \tilde{\phi} .  
\eq
Here again the sign of $c_1$ is determined by the sign of $r$. 
\bq
\delta{\dot \tilde{\phi} } = \left( 8 \beta_s- \frac{gr} {3 \beta_s^3} \right) \delta \beta= c_3 \delta \beta . 
\eq
When $r <0$ and we choose the previous values- $ r = -0.01, g=2, k=-0.1, \delta =-1, k'^2 = 1.01$ we obtain for $f_1(y)$
\bq
\beta_1=0.505726;~~c_1 =-0.00659119; ~~c_2 = 4.09735; ~~\Omega^2=-0.0270064 , 
\eq 
which suggests that this stationary  solution is unstable.

When $r>0$, 
 the sign of $c_3$ depends on whether one is taking the positive or negative  
sign in  Eq. (\ref{posneg}).
For our initial conditions $c_3$ is positive for $f_{2+}$ and negative for $f_{2-}$ corresponding to 
the $\pm$ choice in Eq. (\ref{posneg}). Combining these two equations  we now find:
\bq
 \delta {\ddot  \beta} \mp \Omega^2 \delta \beta =  
\delta {\ddot  \tilde{\phi}} \mp \Omega^2 \delta \tilde{\phi} =0,
\eq
where $\Omega^2 = |c_1 c_3 |>0$.  This suggests that the fixed point associated with $f_{2+}$  is  unstable and $f_{2-}$ stable under small perturbations.
For the solution with r=0.05, for  $f_1$ we find that  
\bq
c_1= .034; ~~ c_2= 4.16915;~~ \Omega = .378701 . 
\eq
For $f_{2+}$
\bq
c_1 = .034;~~ c_3 = 3.58;~~ \Omega = .351073 . 
\eq
For $f_{2-}$
\bq
c_1= .1251; ~~ c_3 = -13.0305;  ~~\Omega = 1.27676 . 
\eq

For the solution with r=0.005, we find for  $f_1$ that   
\bq
c_1= 0.00332219; ~~ c_2= 4.03982;~~ \Omega = 0.115849 . 
\eq
For $f_{2+}$
\bq
c_1 =0.00334441 ;~~ c_3 =3.95982;~~ \Omega =0.115079 , 
\eq
which suggests instability.
For $f_{2-}$
\bq
c_1= 0.0406897; ~~ c_3 = -48.1771;  ~~\Omega = 1.40011 . 
\eq

The small oscillation frequencies for the stable solutions $f_1$ and $f_{2-}$ are borne out by numerical simulations.
However, for the predicted unstable case, at early times,  $0< t < 650$ the solutions 
exhibit a small oscillation with frequency $
\Omega =0.0448$.  After that time the solution switches to the separatrix solution  with large oscillations in $\beta$ and $\phi$ but small oscillations in $p$ and $q$.

\subsection{$\kappa =1$}
For comparison let us review what happened in the case $\kappa =1$ which is quite different. 
   At $\kappa=1$ the Lagrangian is given by:
   \bq
L=-2 ~r~\sqrt{\frac{2}{g}} \pi ~\text{sech}\left(\frac{\pi \tp(t)}
{2
   \beta (t)} \right) \cos (\tphi (t) )+4 \frac{\beta (t)}{g} 
   \left(  \tp(t) {\dot \tq} (t)-\tp(t)^2- \dot {\phi} (t)+\delta-k^{2} + \frac{\beta^2}{3} \right) . 
   \eq
  ~From this we get the following four equations:
  \bq
  \frac{d}{dt} (\beta  \tp ) = 0 ,  \label{tc}
   \eq
    \bq
  \dot{\beta} = - \frac{\pi  r} {2}  \sqrt{2g}  ~\sech C \sin (\tphi (t) ) , \label{td}
   \eq
   \bq
 \dot{\tq} = 2 \tp(t)  -  \frac{\pi ^2 \sqrt{2 g}~ r \tanh C \sech  C
   \cos (\tphi (t))}{4 \beta (t)^2}  , \label{te}
   \eq
   \bq
   \tp(t) {\dot \tq}(t)-\tp(t)^2-\dot{\tphi}(t)+\delta-k^{2} 
+ \beta (t)^2 =   \frac{r~\tp(t)  \pi ^2 \sqrt{2g}~  \tanh C~ \sech C 
   \cos (\tphi (t))}{4 \beta (t)^2} , 
 \label{tf}
   \eq
   where
   \bq
   C = \frac{\pi \tp(t)}{2 \beta(t)} . 
   \eq
~From Eq. (\ref{tc}) we obtain 
   \bq
   \beta(t) \tp(t) = constant = a_1 . 
   \eq
  Combining Eqs. (\ref{te}) and (\ref{tf}) we find 
  \bq
  \dot{\tphi}(t) = \tp^2 -  \frac{r~\tp(t)  \pi ^2 \sqrt{2g}~  \tanh C~ \sech C 
   \cos (\tphi (t))}{2 \beta (t)^2}  + \beta^2 - k'^2 . 
   \eq
  The stationary solutions have 
  \bq
  \dot{\tphi}(t) = \alpha_s;~~ \tp = \tp_s;~~ {\dot \tq} = v_s; ~~\beta = \beta_s; ~~ C_s = \frac{\pi \tp_s}{\beta_s} . 
  \eq
  We have $\tphi = n \pi$  and we can restrict ourselves to $0, \pi$.
  Thus
  \bq
  v_s = 2\tp_s \mp 
   \frac{\sqrt{2g}~ \pi ^2 ~ r~ \tanh C_s  ~\sech C_s }{4 \beta_s^2} , 
   \eq
    \bq
\alpha_s = \tp_s^2  \mp   \frac{r~\tp_s  \pi ^2 \sqrt{2g}~  \tanh C_s~ \sech C_s 
}{2 \beta_s^2}  + \beta_s^2 - k'^2 . 
   \eq
We are interested in small oscillations around the stationary solutions.  We will choose $\beta$ and $\tphi$ as the independent variables with $\tp = a_1/\beta$, and $C =  \frac{\pi a_1}{2 \beta^2(t)}$.   Letting $\beta = \beta_s + \delta \beta$, and $\tphi = \alpha_s t  + \delta \tphi$
we obtain for small oscillations:
\bq
\delta {\dot \beta} = \mp \frac{\pi r}{2} \sqrt{2 g} \sech C_s \delta 
\tilde{\phi} = \mp c_2 \delta \tilde{\phi} , 
\eq

\ba
\delta \dot{ \tphi}&& = \left( 2 \beta_s - \frac{2a_1^2}{\beta_s^3}  \pm r~\pi^2 \sqrt{2 g} \frac{a_1}{2 \beta_s^4} \left[ 3  \tanh C_s~ \sech C_s  + 
\frac{\pi a_1}{\beta_s^2} ~\sech C_s (2 \sech^2 C_s -1) \right]  \right) \delta \beta \nonumber \\
&&= c_3 \delta \beta . 
\ea
   Thus we get the equations
  \bq
  \delta {\ddot \beta} \pm c_2c_3  \delta \beta =0\,, ~~   \delta {\ddot \tphi} \pm c_2c_3 \delta  \tphi=0 .
  \eq
When $C=0$ instead we have
\begin{eqnarray}
\tp&=&0   ;  \dot{\tq} = 0,\nonumber  \\
\dot{\beta} &=& - r \pi \sqrt{g/2} \sin (\tphi), ~~ \dot{\tphi}=    \delta-k^2+  \beta^2  .   \label{tphi1a} // 
\end{eqnarray}
Thus the stationary solution is 
\bq
 \beta_s^2 = k'^2 , 
\eq
And the small oscillation equations are
\bq
\delta{\dot{\beta}} = \mp r \pi \sqrt{g/2} \delta \tphi; ~~ \delta{\dot{\tphi}} =  2 \beta_s \delta \beta .
\eq

\section{Damped and forced NLSE (theory)}

In performing numerical simulations, one is interested in adding damping to the problem that we have studied earlier.
The damped and forced NLSE is represented by
\bq \label{df1}
 i \frac{\partial}{\partial t} \psi + 
\frac{\partial^2}{\partial x^2} \psi + {g }(\psi^\star \psi)^{ \kappa} \psi+
\delta \psi  = r e^{-i( k x+ \theta)}-i \alpha \psi, 
\eq
where $\alpha$ is the dissipation coefficient. 
This equation can be derived by means of a generalization of the Euler-Lagrange 
equation  
\bq \label{df2}
\frac{d}{dt} \frac{\partial {\cal L}}{\partial \psi_{t}^{*}} + 
\frac{d}{dx} \frac{\partial {\cal L}}{\partial \psi_{x}^{*}}-
\frac{\partial {\cal L}}{\partial \psi^{*}}= 
\frac{\partial {\cal F}}{\partial \psi^{*}_{t}},
\eq
where the Lagrangian density reads  
\bq \label{df3}
{\cal L} = \frac{i}{2} (\psi_{t} \psi^{*}-\psi_{t}^{*} \psi)-|\psi_{x}|^{2}+ 
\frac{g} {\kappa+1} (\psi^\star \psi)^{\kappa+1}  +\delta |\psi|^{2}-r e^{-i( k x+ \theta)} \psi^{*}- r e^{i( k x+ \theta)}\psi,
\eq
and the dissipation function is given by 
\bq \label{df4}
{\cal F} = -i \alpha (\psi_{t} \psi^{*}-\psi_{t}^{*} \psi).
\eq
Inserting the ansatz Eq. (\ref{varpsi}) into 
(\ref{df4}) we obtain  
\bq \label{df5}
{\cal F} = -2 \alpha A^{2} f_{v}^{2} (\beta_{v}(t)(x-q(t))) (\dot{p} (q-x)+p 
\dot{q}-\dot{\phi}).
\eq
Integrating this expression over space we obtain 
\bq \label{df6}
F = -2 \alpha C_{0} \frac{A^{2}}{\beta} (p \dot{q}-\dot{\phi}).
\eq
On the other hand   $L = \int dx  {\mathcal L} $ is given by Eq.(\ref{efflag}).
Since $F$ contains $\dot{q}$ 
and $\dot{\phi}$, only the equations for $\beta$ and $p$ could be changed 
by the damping term. For $\kappa=1$, we know that the damping  
only  affects the equation for $\beta$ \cite{mqb}. For $\kappa=1/2$ 
we have the same scenario, i.e. now the equation for $\beta$ reads
\bq \label{df7}
\dot{\beta} = -\frac{2}{3} \alpha \beta -\frac{g r C}{6 \beta} 
\frac{\sin(B)}{\sinh(C)}. 
\eq 
The factor $2/3$ comes from $2/(2/\kappa-1)$.

\section{Numerical Simulations on the unforced and the forced NLSE}

In this section we would like to accomplish three things.  
First, we would like to show that our 
numerical scheme as applied to exact soliton solutions with either 
$r=0$ or $r \neq 0$ leads to 
known results ($r=0$) and to new results that the exact solutions to 
the forced problem are metastable 
and at late times become a solitary wave whose parameters oscillate in time.  
Secondly, we want to compare the results of solving the four CC equations 
for the variational  parameters $\beta$, $q$, $p$, $\phi$ with a numerical determination of these quantities found by solving the FNLSE.  We will find that if $r$ is small compared with the amplitude of the solitary wave, that solution of the CC equations gives a good description of the wave function of the FNLSE.  Finally, we would like to demonstrate that the regions of  stability for the  solitary waves of the FNLSE are well determined by studying the phase portrait as well as the  $p(v)$ curve for the approximate solution found by solving the CC equations.  Most of the simulations will be restricted to $\kappa = 1/2$ except for a discussion of blowup of solutions for $\kappa \geq 2. $

\subsection{Numerical methodology} 
Before displaying the results of our simulations, we would like to say a little bit about the numerical method we used and the boundary conditions, since we are constrained to perform the calculation in a box of length $2 L$.  We have allowed the length to vary from 100 to 
400. The number of points used on the spacial grid was $2L/ \Delta x$.  
The numerical simulations were performed using a
4th  order Runge-Kutta method. $N+1$ points are used on the spatial grid $n=0,1, \ldots N$.
When studying the  exact solutions we have used  three different boundary conditions.  For $r=0$ we use 
``hard wall" boundary conditions, where the wave function vanishes at the boundary:
\bq
\psi(\pm L,t)=0.
\eq
For the case of $r \neq 0$ and only when studying the exact solutions we have used mixed boundary 
conditions (see (\ref{mixed})). Otherwise, we have used periodic  boundary conditions
\bq
 \psi(-L,t)=\psi(L,t),~~\psi_{x}(-L,t)=\psi_{x}(L,t).
 \eq
In one of our simulations we have compared the use of periodic vs. mixed boundary conditions and found that the differences in the evolution of both 
$\beta(t)$ and $\psi(x,t)$ are hardly visible by the ``eyeball method"  (see Fig.\ref{s2d})
The other parameters related with the 
discretization of the system are increasing values of $L$, namely   
$L=50$, $L=62.8$, 
$L=100$ or $L=200$.  We have chosen as our grids in $x$ and $t$, 
$\Delta x= 0.05$ and $\Delta t=0.0001$, (such that $\Delta t < (1/2) (\Delta x)^{2}$).

We next need to determine the parameters $q$ and $\beta$ used in the CC equations.
We determine $q(t) $ from our numerical simulation by equating it to the  
value of  $x$ for which 
the density of the norm $|\psi|^2$ is maximum. 

To determine $\beta(t)$ for finite energy solitary waves,  we assume that the variational parameterization of $\psi(x,t)$, namely Eq. (\ref{ansatz})
with $A$ given by Eq. (\ref{choicea}) is an adequate description of the wave function.
If we do this and determine $\psi(x=q(t))$ then we have that 
\bq
|\psi|^2 (x=q(t)) = A^2 = \left[\frac{\beta^2 (\kappa+1)}{g \kappa^2} \right]^{1/\kappa}.
\eq
Then we determine $\beta$ as follows: 
\bq
\beta=\sqrt{[g \kappa^2 |\psi|^{2 \kappa} (x=q(t))]/(\kappa+1)}. 
\eq
In the particular case when we are studying the time evolution of the  exact solution for $\kappa=1/2$, 
i.e. Eq. (\ref{psi1}), with conditions of 
Eq. (\ref{re2}), we need to subtract off the constant term to determine $\beta(t)$.  From Eq. (\ref{psi1}) we can obtain  $\beta(t)$ from
\bq
\sqrt{|\psi|^2 }_{|x=q(t)} =a + \frac{6 \beta(t)^2}{g} .  \label{exactbeta}
\eq
  One can also compute the momentum $P(t)$ given 
by  Eq. (\ref{momentum}).

As an initial condition in most of our simulations (when we are not discussing the exact solution), we will use an approximate solution 
given by the variational ansatz Eq. (\ref{ansatz}) with initial conditions
 for $\beta_{0}$, $p_{0}$, $q_{0}=0$ and $\phi_{0}$.  In comparing with the CC equations, we will solve 
the four ordinary differential equations for $\beta$, $p$, $q$, $\phi$  which for arbitrary $\kappa$ are 
given in Eq. (\ref{rela3}) to Eq. (\ref{dottphi1}).

\subsection{Numerical simulations of PDE for $r=0$, arbitrary $\kappa$ }

The initial conditions are those of the  exact 1-soliton solution of the 
unforced NLSE given 
by (\ref{exact}). 
For $r=0$ and $\delta =0$, the stability of the NLSE has been well studied as a function of $\kappa$.  For $\kappa<2$ the solutions are known to be stable, and for $\kappa > 2 $ the solutions are unstable.  For $\kappa =2$ there is a 
critical mass above which the solutions are unstable.  The nature of the solution when it is unstable has been studied in variational approximations  \cite{cooper3} as well 
as  using various numerical  algorithms  \cite{numerical, kevrekedis}.   Here we want to show that our code reproduces these well known facts.    We are also interested  in the effect of $\delta$ in our simulations and we find that the critical value of $\kappa$ is independent of $\delta$.
This agrees with our arguments earlier on the effect of scale transformations on the stability of the exact solutions for $r=0$. That is, stability does not depend on $\delta$. 

\subsection{Simulations at $r=0$, different values of $\kappa$}
In this section  we study the numerical  stability of the exact solutions for  $r=0$ for different values 
of $\kappa$ ($\kappa \in[0.25, 3]$).  First we  set $r=0$, $\delta=0$, $g=2$ in NLSE 
and we start from the exact soliton solution (\ref{exact})  with 
$\beta=1$, $p=0.01$ and $\phi_0=0$. In this simulation the boundary conditions chosen were $\Psi(\pm L,t)=0$.  We notice for
$\kappa <2$ as shown in Fig. (\ref{s1}) the solitary wave moves to the right and maintains its shape.  

Indeed, in the simulations the solitons are not stable for $\kappa \ge 2$ 
(see Fig.\ \ref{s2}), the amplitude of the unstable soliton 
grows and the soliton becomes narrow. Also the soliton moves only slowly to the right while blowing up at a finite time.
 In Fig. \ref{fs2a} upper panel  we show the real and imaginary 
parts of the field for $\kappa=1.5, 2, 3$ for the final time 
of the simulations.  The result of adding a term proportional to $\delta$ does not alter the behavior of the solitary waves. 
 Choosing  $\delta=-1$ we obtain similar results as for $\delta =0$, i.e. the soliton is unstable for 
$\kappa \ge 2$. In Fig.\ \ref{fs2a} lower panel  we show the evolution of $\beta$ for short 
times and $\kappa=1.5; 2; 3$. This shows once we reach the metastable solution for $\kappa =2$, we see $\beta(t) \rightarrow \infty$ in a finite amount of time.  This agrees with our analysis based on Derrick's theorem (see Eq.(\ref{Derricktheorem}.)), and with the discussion of the critical mass needed for blowup for $\kappa=2$  in \cite{cooper3}.
 For $\delta=-1$ we have also investigated the constant phase solutions 
that fulfill the condition (3.5). We have studied numerically the case  $r=0$, 
$\delta=-1$, $g=2$ with initial conditions $p=0.01$, $\phi_0=0$ and 
$\beta=\sqrt{-\kappa^2 (\delta+p^2)}$ and varied $\kappa \in [0.25, 3]$. 
We found that again  the soliton becomes unstable for $\kappa \ge 2$ in accord with our result that instability as a function of $\kappa$ does not depend on $\delta$.  
Our numerical experiments for the case $r=0$  show that our codes reproduce well known results for the stability of the solutions. 
In a future paper we will compare the numerical solutions in the unstable regime with the predictions of the CC method.  Here we will use the exact form of the solution rather than the post-Gaussian trial functions we used earlier \cite{cooper3} as well as include an additional variational parameter in the phase that is canonically conjugate to the width parameter $\beta$.
To estimate the finite size effect on the definition of $\beta$ we notice in Fig. \ref{s2c}, that the values of $\beta(t)$ in the simulation of the 
NLSE deviate from the exact constant value of $0.5$ by approximately 
$0.25 \%$.  By increasing $L$ one could reduce this error.

\subsection{On the exact solutions for $r \ne 0$ and $\kappa=1/2$}
Earlier we showed that for $r \ne 0$ and $\kappa=1/2$ there are exact solutions of the form (see Eq. (\ref{fy3}))
\ba \label{fy3a}
f(y)&& = - {\rm sign}(r) \left[ {a}+ {b} \sech^2 \beta(y,t) \right] , \nonumber \\
\ea
with $a$, $\beta$ and $b$ given by Eq. (\ref{re2}).
 In the numerical simulation shown in  Fig. \ref{s2c}  and  Fig. \ref{s2cc}   the parameters used in the simulation are  $\kappa=1/2, \delta = -1, g=2, k=0.1, r=-0.075$.  For that case:
\bq
f(y) = 0.486113 \text{sech}^2(0.402539 y)+0.0904622.    \label{reffigs2cc}
\eq
  The unstable stationary variational solution corresponding to this  exact solution is 
\bq 
f_{2+}(y) =0.745535 \text{sech}^2(0.498509 y).
\eq
These solutions are shown in Fig. \ref{ugga}.  As we can see from the numerical solution of the FNLSE shown in Fig. \ref{s2c},   the exact solitary wave oscillates in amplitude and width which shows that the original exact solution was unstable.  In spite of this the solitary wave neither dissipates nor 
blows up.  The  width parameter oscillates from its initial value of around $0.4$ to an upper value of around $0.8$  with a period $T$ of around
$T=20$.  If one reduces the driving terms to $k=0.01$ and $r=-0.005$, the solitary wave again oscillates in amplitude and width but the amplitude of the oscillation is only $10 \%$ of the total amplitude  and the oscillation frequency is reduced from the previous case as is shown in Fig. \ref{s2ccc}.

In Fig. \ref{s2d} we compare the behavior of $\beta(t)$, and the late form of the wave functions using two types of boundary conditions (BCs):
periodic   BCs  ($\psi(-L,t)=\psi(L,t)$, $\psi_{x}(-L,t)=\psi_{x}(L,t)$) shown as dashed lines, and ``mixed" BCs  which are shown as solid lines.  In this simulation the two boundary conditions lead to almost identical results.  Most of the simulations (except the ones starting from the exact solution) were performed using periodic boundary conditions.
Next we look at the case where $r>0$ where for amplitudes $A(t) >0$ there are no exact solutions.  However, the stationary solutions 
of the CC equations that are stable to linear perturbations do lead to solitary waves that have widths whose oscillations have only very small amplitude.  To show that we start with the stationary solutions with different positive $r$ but other parameters $(\kappa, \delta, g, k)$ the same as in Fig. \ref{s2c}.   We show this in Fig. \ref{s2bpr}  where we consider solitary waves moving to the left initially with constant velocity $v=-2k$.  We see that as we increase $r$ the amplitude of the solitary wave increases and in Fig. {\ref{s2cpr} we see  that the average value of $\beta$ also increases with $r$. 
We notice in Fig. \ref{s2cpr}  that the oscillations in $\beta$ get more chaotic as we increase the value of $r$.

\subsection{Comparison of numerical simulations of the FNLSE with the solution of the equations for the collective coordinates}
In this section we would like to show that the solution to the CC equations gives quantitatively good results for the collective variables $p(t), \beta(t),
q(t)$ and $\phi(t)$ when these quantities are calculated directly from the solution of the FNLSE.  We also want to show that the criterion for stability of the CC equations, namely a study of the phase portrait or the $p(v)$ curve,  gives an accurate measure of what initial values of collective variables lead to stable vs. unstable solitary wave solutions.   First we discuss the linearly stable stationary solution Eq. (\ref{f1c}) that is shown in Fig. \ref{f1y}.  
In the CC equation $\beta$ remains at the fixed value $\beta(0) =\beta_{s}=0.51773$. 
Using this as an initial condition in the FNLSE solved using periodic boundary 
conditions, one finds that the solitary wave 
has a mean value of $\beta$ only 2 \% different from the solution, Fig. \ref{fig18A}. 
For $0<t\le30$, phonons (short for linear excitations) are radiated by the soliton which travel faster than 
the soliton. Coming from a boundary these phonons reappear and collide with the soliton producing 
the increased oscillations seen in  Fig. \ref{fig18A}. For the system of 
length $2 L=125.6$, these increased oscillations begin at $t_{c} \approx 125$. 
When the system size is doubled, both the soliton and the phonons have to cover 
doubled distances before the collisions begin at $t_{c} \approx 250$.

 If we add a little bit of damping and start from a value of $\beta_0$  which dissipates to a stationary point then all the collective variables are well approximated by 
the solution of the CC equation as shown in the middle panel of Fig. \ref{fig1}.  Here we have chosen   $g=2$, $k= 0.1$, $r=0.05$,
 $\delta=-1$, 
$\theta=0$, $\alpha=0.05$ .  For these values,  the stationary solution is approximately given by  Eq. (\ref{mugga1}) which has a value of $\beta_s = 0.50512$. 
 In the middle panel of  Fig. \ref{fig1} we use the  initial condition
$\beta_0 = 1.$  We find both the CC equations and the solution of the FNLSE 
relax to the approximate stationary solution of Eq. (\ref{mugga1}). 

In Fig. \ref{fig7} we turn off the damping and see how both the simulation and 
CC equations evolve. 
 Solving the CC equations with the initial condition $\beta_0 = 1$,  
one finds from both the orbit of this initial condition in the phase portrait 
(see left bottom panel) and the $p(v)$ curve that the solitary wave should be 
stable.  In the middle panel we see that both the exact solution and the 
solution to the CC equation for $\beta$ oscillate with 
1\% oscillations  around either the initial value $\beta_0 = 1$ 
(for the numerical solution) and a slightly shifted value ($0.994$) 
for the CC equation.  The actual solution has another oscillation 
frequency for the height at maximum. 
 
 In Fig. \ref{fig19A}  we study the time evolution of a stationary solution that 
is known to be unstable.  We use the of solution Eq.(\ref{f2}).
  Both the CC equations and the simulation of the FNLSE show that this solitary wave develops into a solution that is represented by a separatrix in the phase portrait. The negative slope of the $p(v)$ curve also predicts instability. 
 
The soliton stability depends strongly on $\beta_0$. For the parameters and initial conditions 
of Fig. \ref{fig9pbc} both stability criteria predict the following pattern: 
instability for $\beta_0 \le 0.484511$, stability for 
$0.484511< \beta_0 \le 0.5458$, instability for $0.5458< \beta_0 \le 0.65$ and 
stability for $\beta_0 \ge 0.66$. In Fig. \ref{fig9pbc} we present two examples: 
a stable soliton ($\beta_0=0.53$) and an unstable one ($\beta_0=0.65$), both confirmed 
by our simulations. We  note, however, 
that the above boundaries between stable and unstable regions do not always fully agree with  the 
simulations: the errors vary between $1.7\%$ and $12 \%$. 

%

  In Fig. \ref{fig12} we show a particular case where the time evolution of an initial condition which was close to an unstable stationary solution of the CC equation exhibits intermittency in both the inverse width (and thus the amplitude $A$  also) as well as  the energy.
  We have also observed intermittency in the solutions of the CC equations for $\kappa=1$. 
\begin{figure}[ht!]
\begin{center}
\begin{tabular}{cc}
\ & \\
\includegraphics[width=7.0cm]{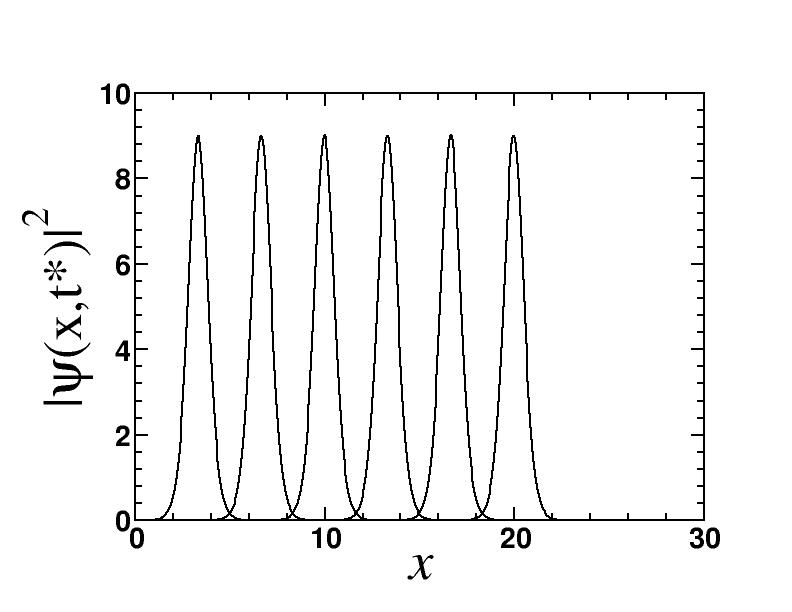}  & 
\quad \includegraphics[width=7.0cm]{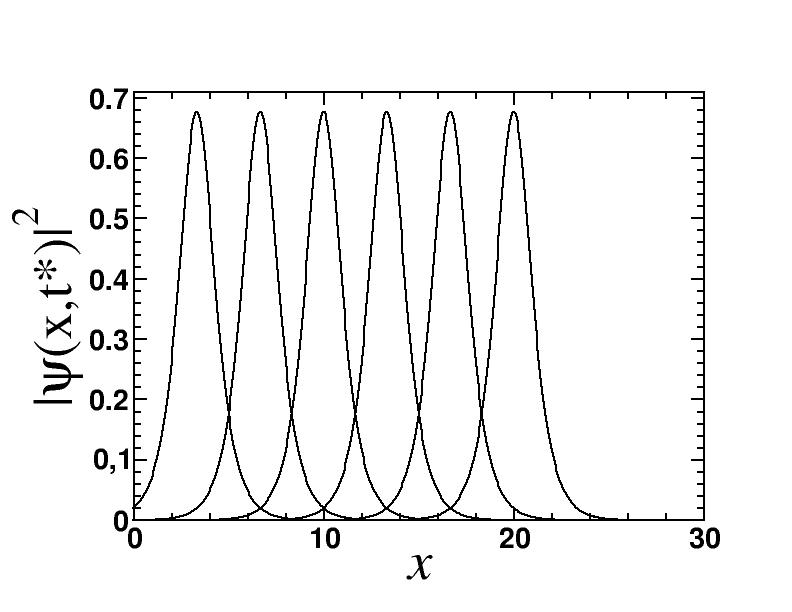}  
\end{tabular}
\end{center}
\caption{$r=0$. Soliton moving to the right at $t^{*}=166.6,\,  
333.3,\, 500,\, 666.6,\, 833.3,\, 1000$. 
Left and right  panel: $\kappa=0.5$ and $\kappa=1.5$, respectively. 
Parameters: $\delta=0$, $g=2$ with initial conditions 
$\beta=1$, $p=0.01$ and $\phi_0=0$.}
\label{s1} 
\end{figure}
\begin{figure}[ht!]
\begin{center}
\begin{tabular}{cc}
\ & \\
\includegraphics[width=7.0cm]{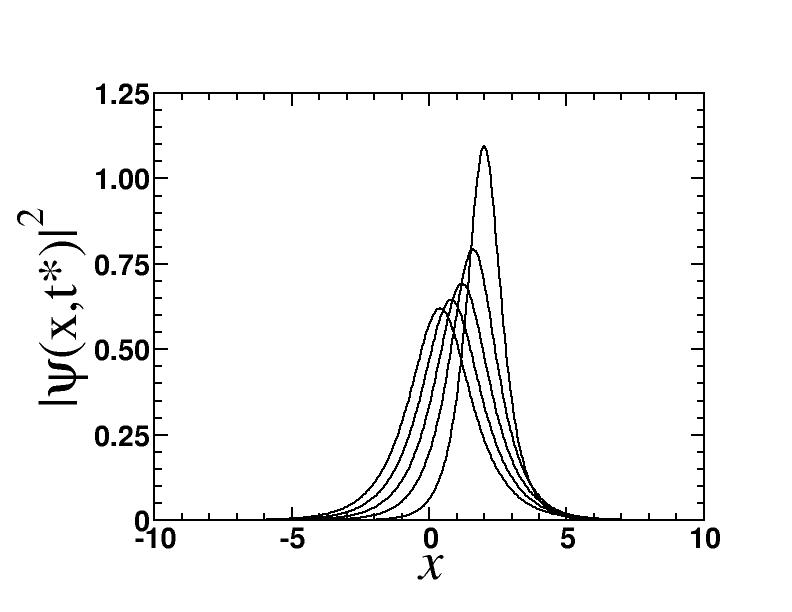}  & 
\quad \includegraphics[width=7.0cm]{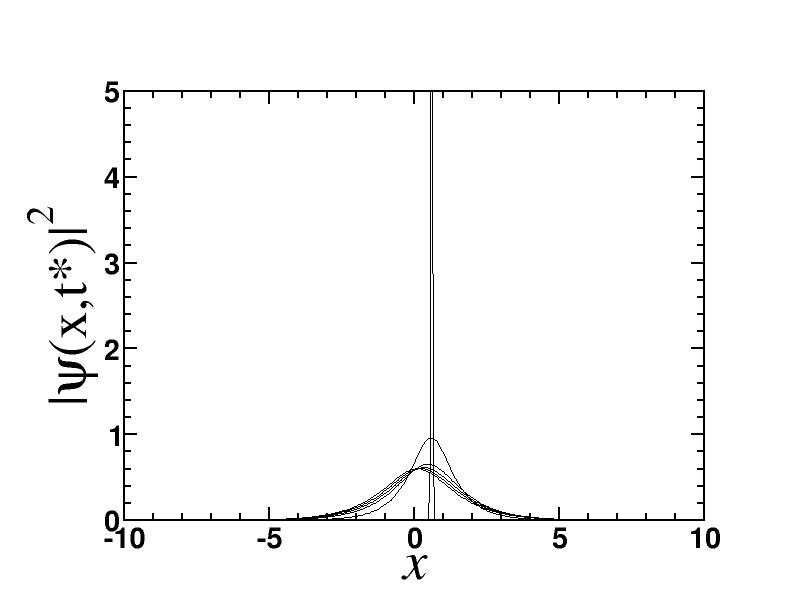}  
\end{tabular}
\end{center}
\caption{$r=0$. Solitary wave moving to the right (blowup when $\kappa \geq 2$). 
Left  panel:  $\kappa=2.0$ 
 (at $t^{*}=166.6;\, 
333.3,\, 500,\, 666.6,\, 833.3,\, 1000$), respectively. 
Right  panel: $\kappa=2.25$  
(at $t^{*}=5.8$; $11.6$, $17.5$, $23.3$, $29.2$, $35.0$).  
Parameters: $\delta=0$, $g=2$ with initial conditions
$\beta=1$, $p=0.01$ and $\phi_0=0$.}
\label{s2} 
\end{figure}
\begin{figure}[ht!]
\begin{center}
\begin{tabular}{cc}
\ & \\
\includegraphics[width=7.0cm]{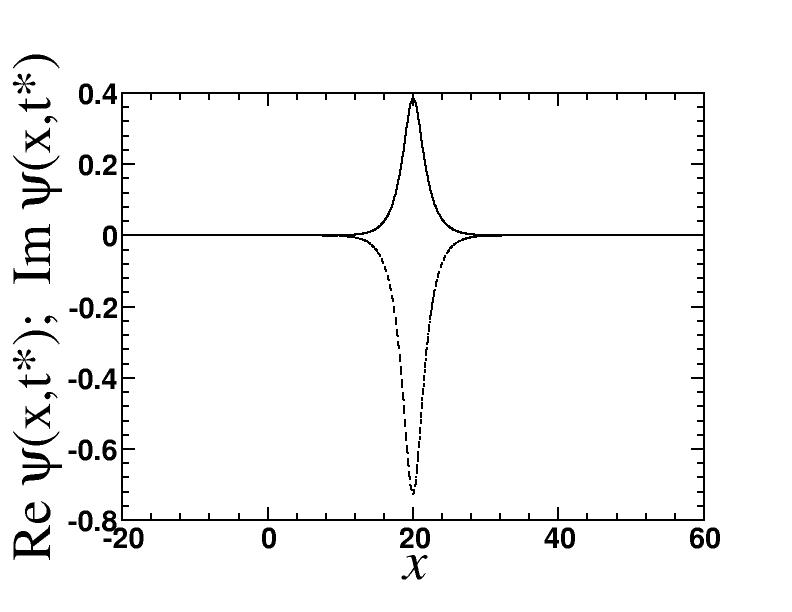}  & 
\quad \includegraphics[width=7.0cm]{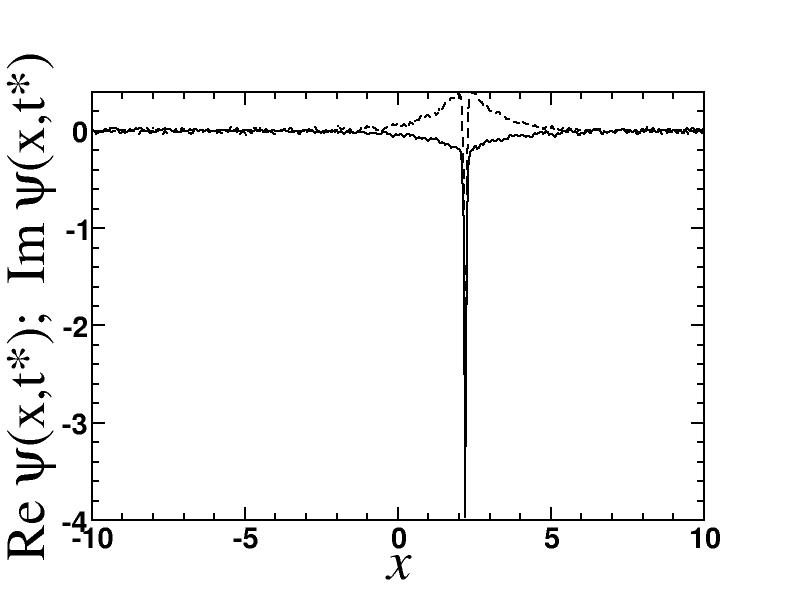}  
 \\
 \ & \\
 \ & \\
 \includegraphics[width=7.0cm]{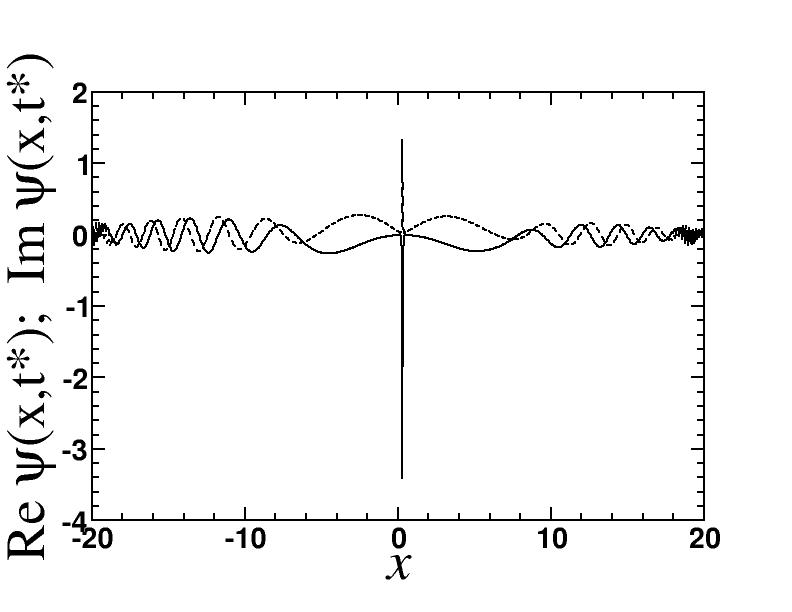}  & 
\includegraphics[width=7.0cm]{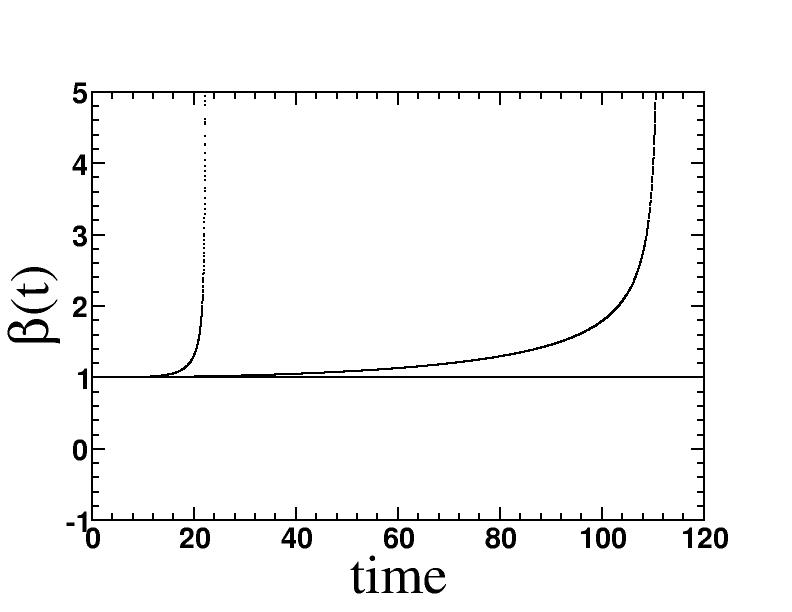}
\end{tabular}
\end{center} 
\caption{$r=0$. Real (solid line) and imaginary (dashed line) parts of the fields for 
$\kappa=1.5$ for $t^{*}=1000$ (left upper panel); 
$\kappa=2.0$ for $t^{*}=120$ (right upper panel);
$\kappa=3.0$ for $t^{*}=25$ (left lower panel). 
 Evolution of $\beta$ from the direct simulations of NLSE. $\kappa=1.5$ 
(solid line); 
$\kappa=2$ (dashed line); $\kappa=3$ (dotted line) (right lower panel).
Parameters: $\delta=0$, $g=2$ with initial conditions
$\beta=1$, $p=0.01$ and $\phi_0=0$.}
\label{fs2a} 
\end{figure}
\begin{figure}[ht!]
\begin{center}
\begin{tabular}{cc}
\ & \\
\includegraphics[width=7.0cm]{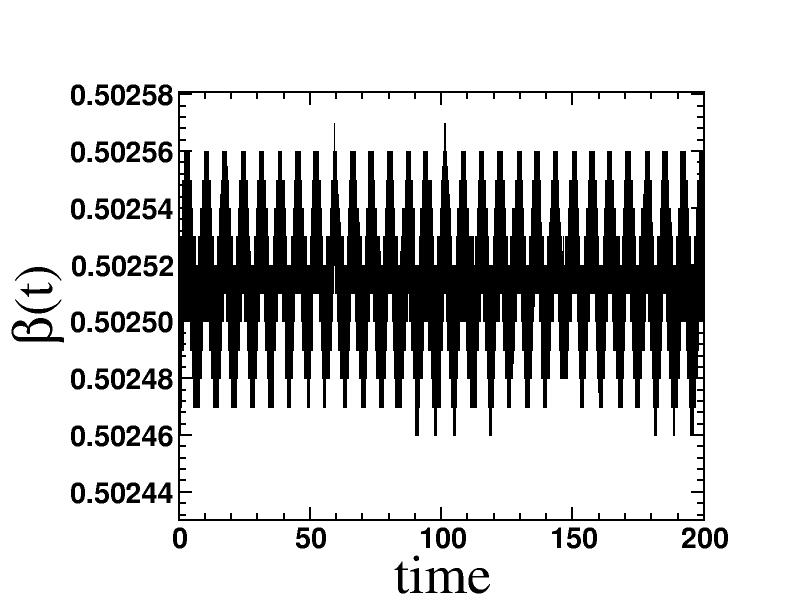}  & 
\quad \includegraphics[width=7.0cm]{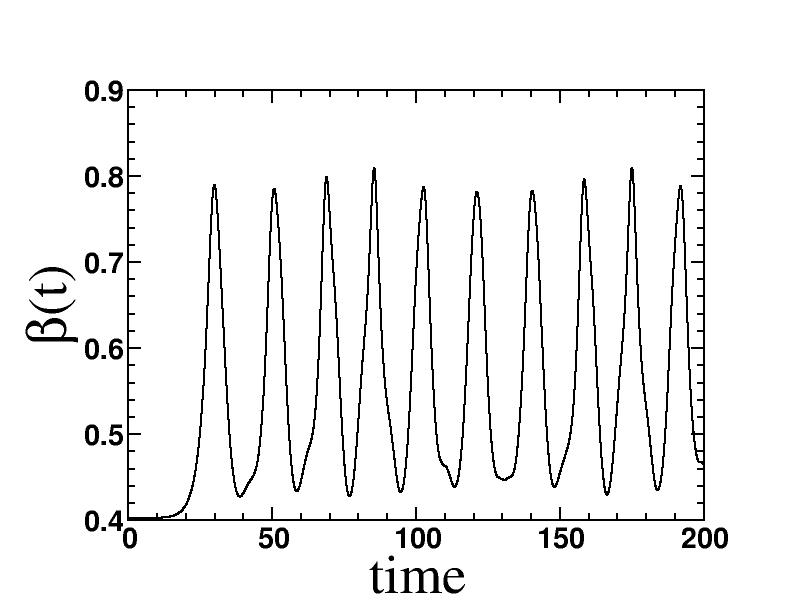}  
\end{tabular}
\end{center} 
\caption{Test of the exact solution Eq.\ (\ref{psi1}) (with the parameters $a$, 
$\beta$ and $b$ determined by Eq.\ (\ref{re1}))   of the perturbed NLSE for 
$\kappa=1/2$, $\delta = -1$, $g=2$, $k=0.1$. 
Here we display the 
inverse width  parameter  $\beta(t)$   of the soliton. 
Left  panel: simulation of the unperturbed NLSE with  $r=0$ ($a=0$) 
corresponding to the constant phase solution of Eq. (\ref{constantphase1}),  
$\beta^2 = - \delta/4$, 
Right  panel: simulation of the perturbed NLSE with  $r=-0.075$.
}
\label{s2c} 
\end{figure}
\begin{figure}[ht!]
\begin{center}
\begin{tabular}{cc}
\ & \\
\includegraphics[width=7.0cm]{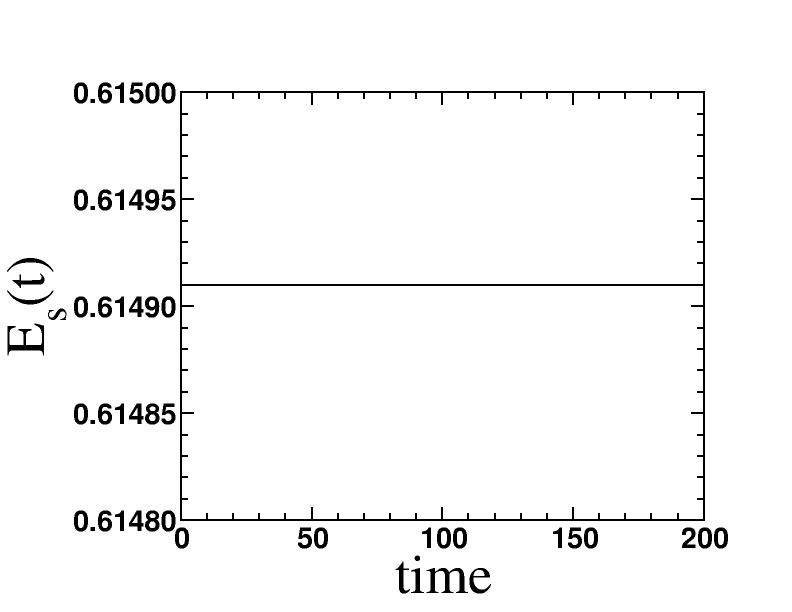}  & 
\quad \includegraphics[width=7.0cm]{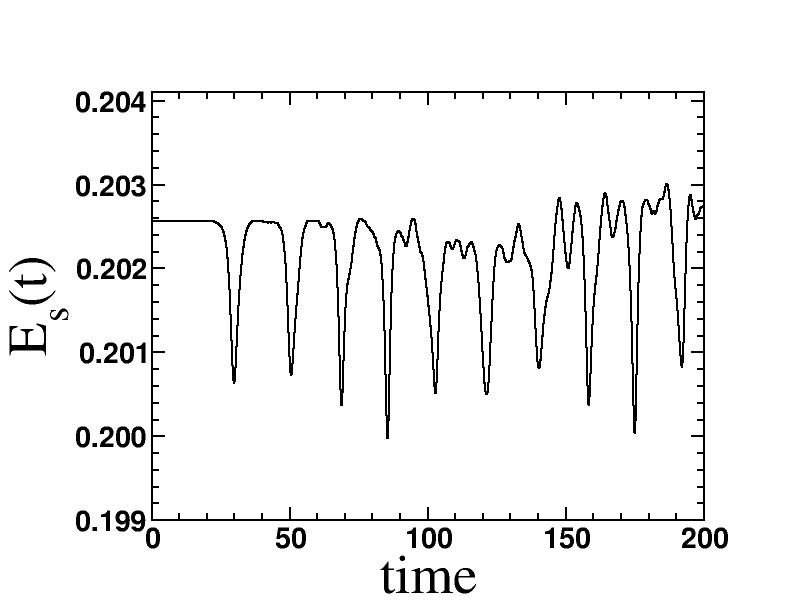}  
\end{tabular}
\end{center} 
\caption{Soliton energy computed by subtracting 
the  background energy density  from the 
total  energy density.  
 Left panel: $r=0$ ($a=0$). 
 Right panel: $r=-0.075$.
Parameters: are the same as in Fig. \ref{s2c}. }
\label{s2cc} 
\end{figure}
\begin{figure}[ht!]
\begin{center}
\begin{tabular}{cc}
\ & \\
\includegraphics[width=7.0cm]{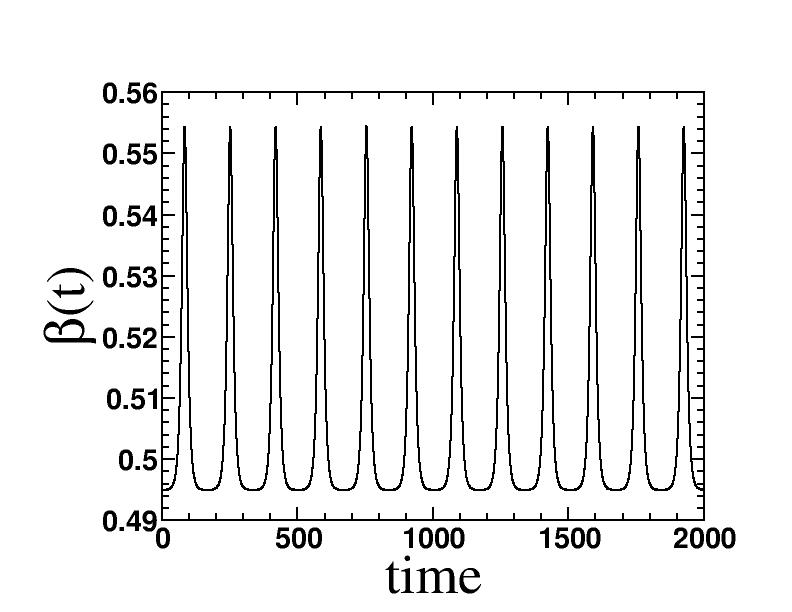}  & 
\quad \includegraphics[width=7.0cm]{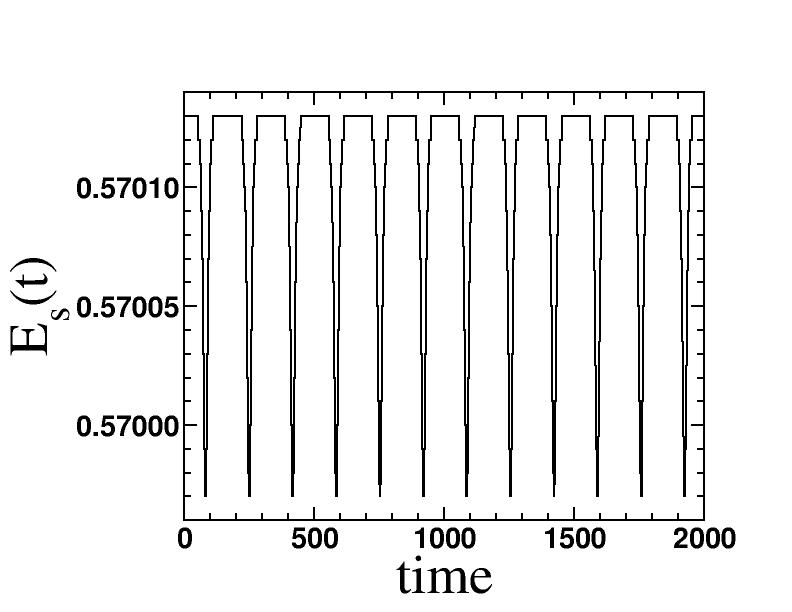}  
\end{tabular}
\end{center} 
\caption{Time evolution of the exact solution of the FNLSE as in Figs.\ 
\ref{s2c} and \ref{s2cc} but using smaller value of the driving terms: 
$k=0.01$ and $r=-0.005$. }
\label{s2ccc} 
\end{figure}
\begin{figure}[ht!]
\begin{center}
\begin{tabular}{cc}
\ & \\
\includegraphics[width=7.0cm]{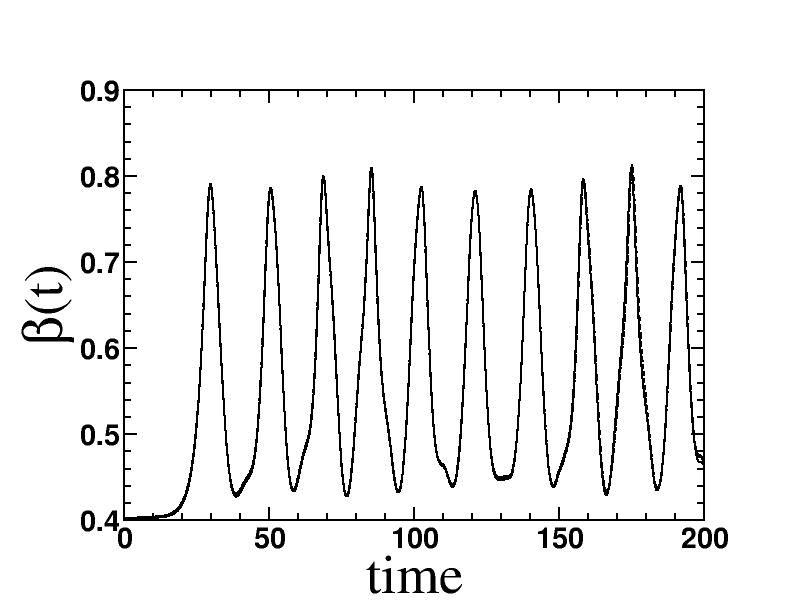}  & 
\quad \includegraphics[width=7.0cm]{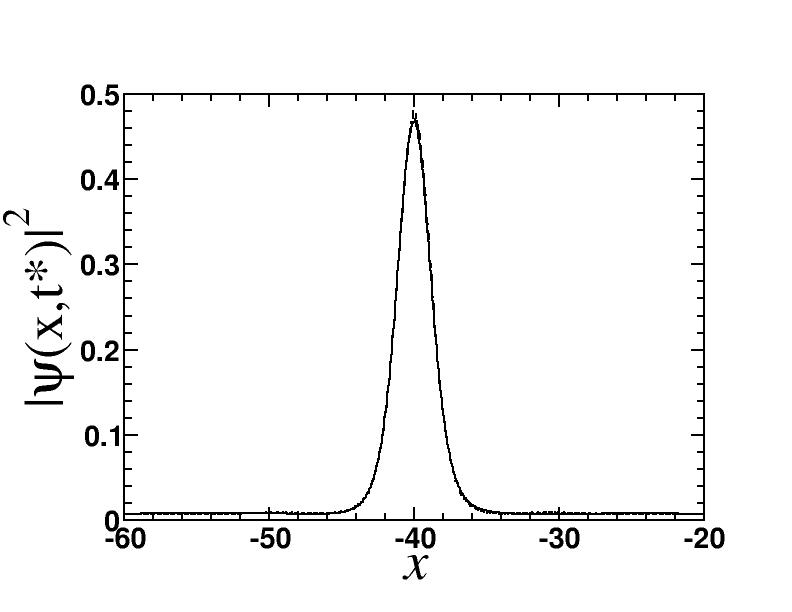} 
\end{tabular}
\end{center} 
\caption{Comparison of the two boundary conditions: periodic (dashed line) vs mixed (solid line)   $r=-0.075$. 
Left panel: $\beta(t)$, 
right panel: soliton profile at $t^{*}=200$.
The parameters are the same as used in Fig. \ref{s2c} and Fig. \ref{s2cc}.
}
\label{s2d} 
\end{figure}
\begin{figure}[ht!]
\begin{center}
\begin{tabular}{ccc}
\ & \\
\includegraphics[width=5.0cm]{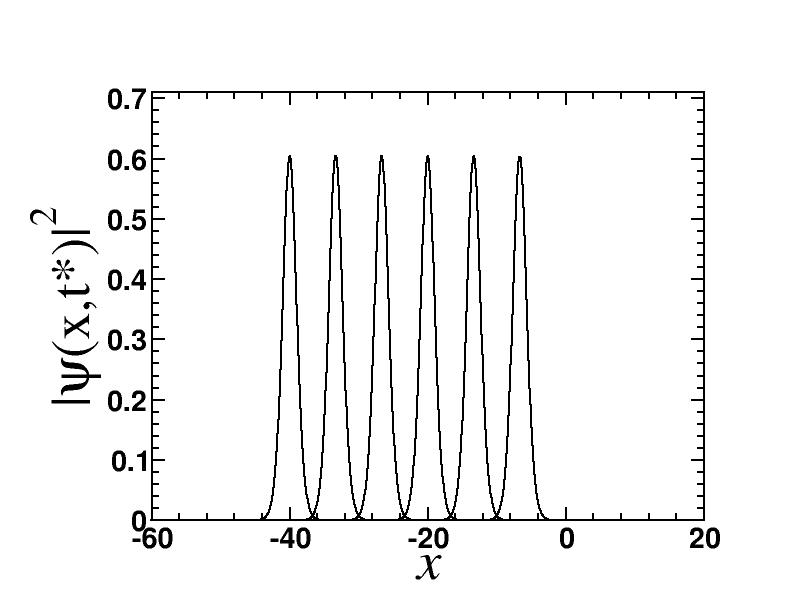}  & 
\quad \includegraphics[width=5.0cm]{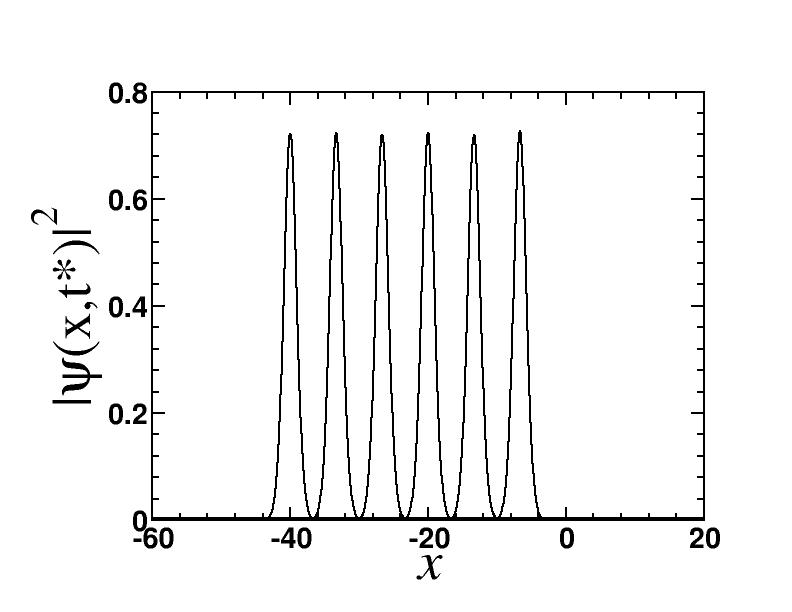}  &
\quad
 \includegraphics[width=5.0cm]{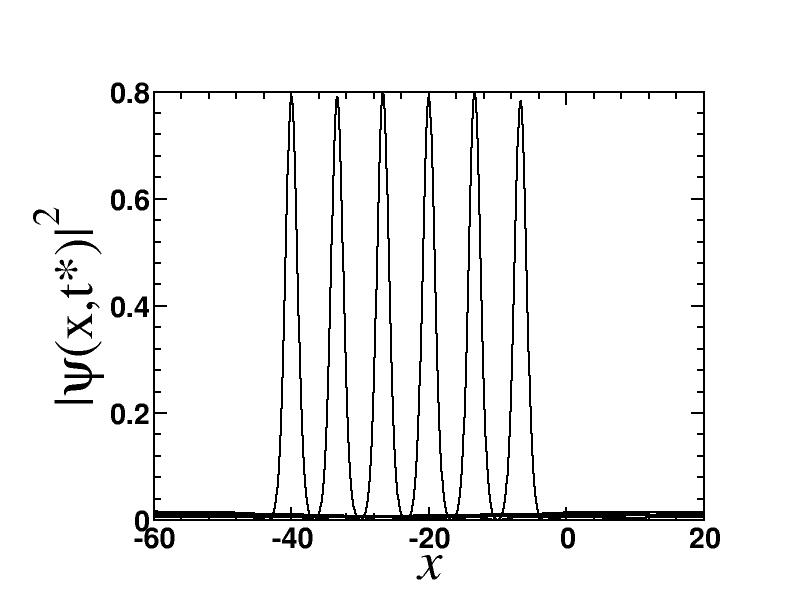} 
\end{tabular}
\end{center} 
\caption{Here we compare the time evolution of a solitary wave traveling to the left initially with constant velocity $-2k$  starting from an approximate stable stationary solution for three different values of positive $r$. The times are 
$t^{*}=33.3$, $66.6$, $100$, $133.3$, $166.6$, $200$. 
We use mixed boundary conditions. 
Left panel: $r=0.01$, 
middle panel: $r=0.05$, 
right panel: $r=0.075$.
Parameters: $\kappa=1/2$, $\delta=-1$, $g=2$, $k=0.1$. Initial condition (\ref{psi1}) with 
(\ref{re1}).} 
\label{s2bpr}
\end{figure}

\begin{figure}[ht!]
\begin{center}
\begin{tabular}{ccc}
\ & \\
\includegraphics[width=5.0cm]{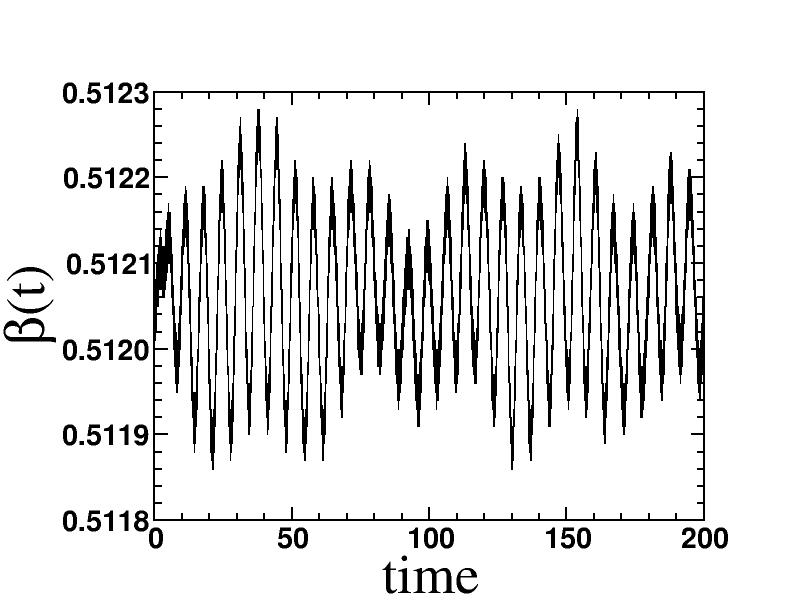}  & 
\quad \includegraphics[width=5.0cm]{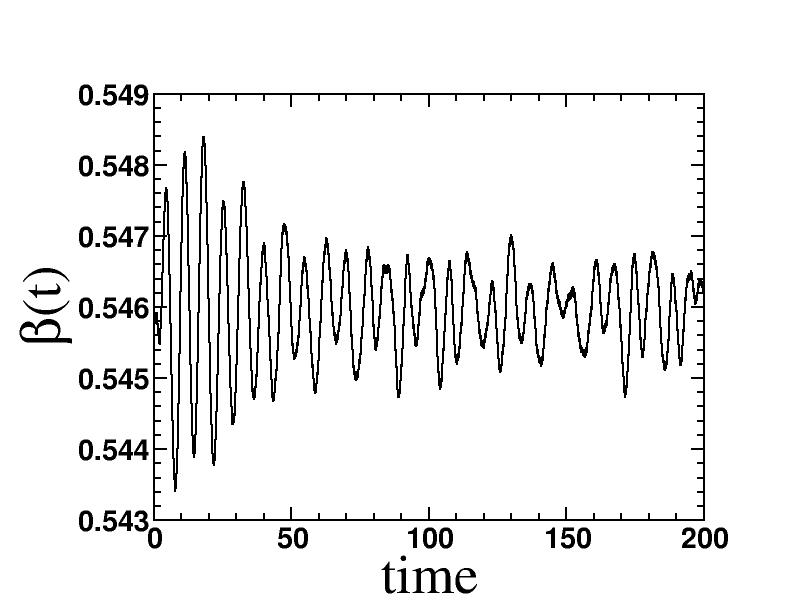}  & 
\quad \includegraphics[width=5.0cm]{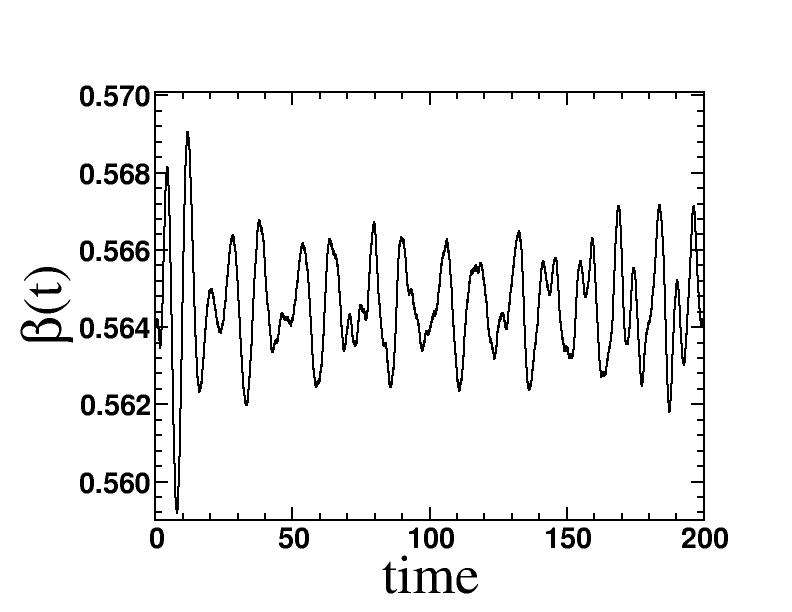}     \\
\end{tabular}
\end{center} 
\caption{Time evolution of the inverse width parameter $\beta(t)$.  
Here we plot $\beta(t)$ for the same initial conditions as in Fig. \ref{s2bpr}. 
Again we have left panel $r=0.01$, 
middle panel $r=0.05$ and  
right panel $r=0.075$.
}
\label{s2cpr} 
\end{figure}

\begin{figure}[ht!]
 \begin{center}
 \begin{tabular}{c}
 \includegraphics[width=7.0cm]{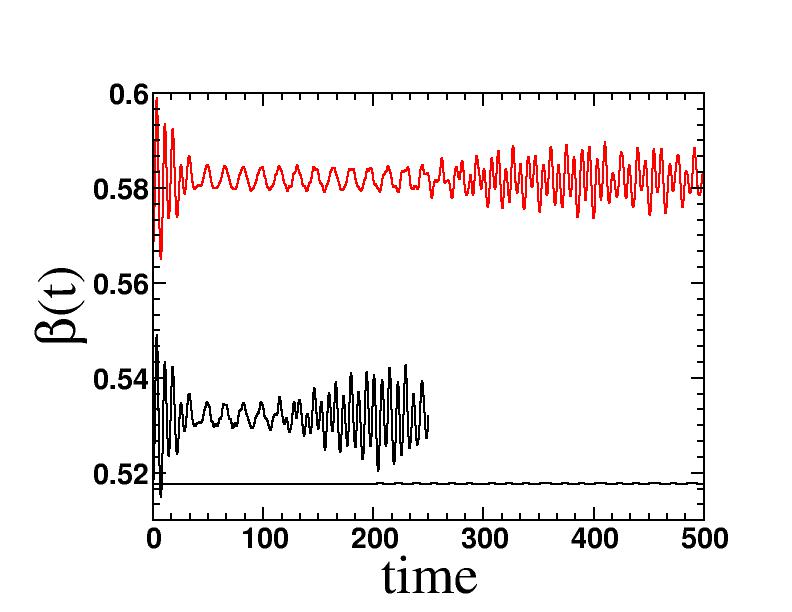}  
 \end{tabular}
 \end{center}
 \caption{Test of stable stationary solution of the CC equations (horizontal line) 
by a simulation of the NLSE with periodic boundary
 conditions (black and red solid lines for system sizes 
$125.6$ and $251.2$, respectively, the latter line is shifted upwards by $0.05$). 
The influence of linear excitations, which are radiated at early times, on the 
soliton oscillations is discussed in the text. The parameters used are 
$\kappa=1/2$, $r=0.05$, $k = -0.1$, 
$\theta =0$, $g=2$, $\delta = -1$, $\alpha=0$, with
 initial conditions $p_0 = -k$, $q_0=0$, $\phi_0=0$ and 
$\beta_0 = \beta_s=0.51773$. For the CC equations we choose 
$p_0 = -k+10^{-5}$ to avoid numerical singularities.  The approximate solitary 
wave is described by Eq. (\ref{f1c}), and is shown in Fig. \ref{f1y}.
 }
 \label{fig18A} 
 \end{figure}
\vspace*{0.5cm} 
\begin{figure}[ht!]
\begin{center}
\begin{tabular}{cc}
\ & \\
\includegraphics[width=7.0cm]{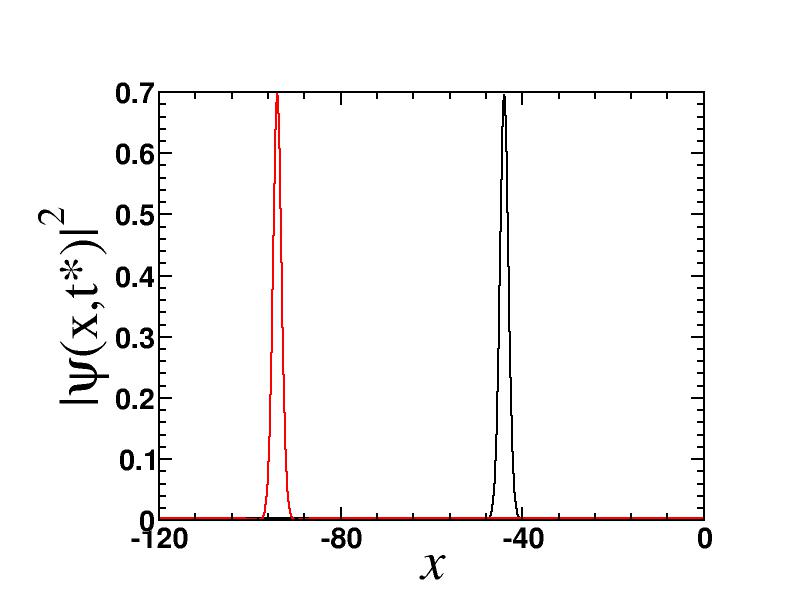}  & 
\quad \includegraphics[width=7.0cm]{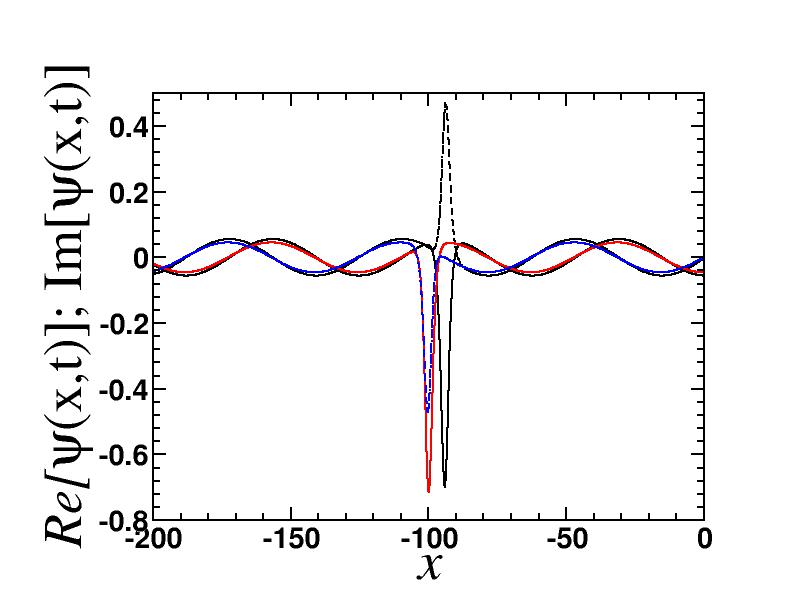}  
\\
\ & \\
\ & \\
\includegraphics[width=7.0cm]{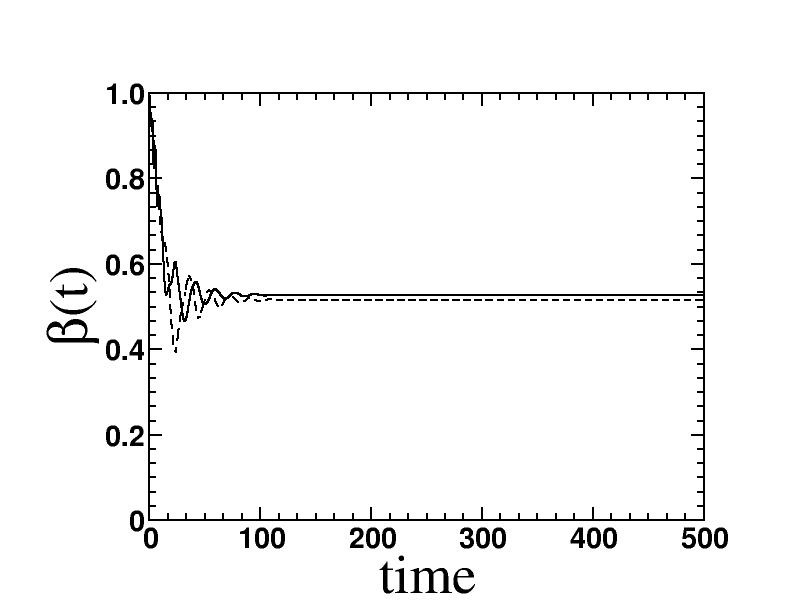}  & 
\quad \includegraphics[width=7.0cm]{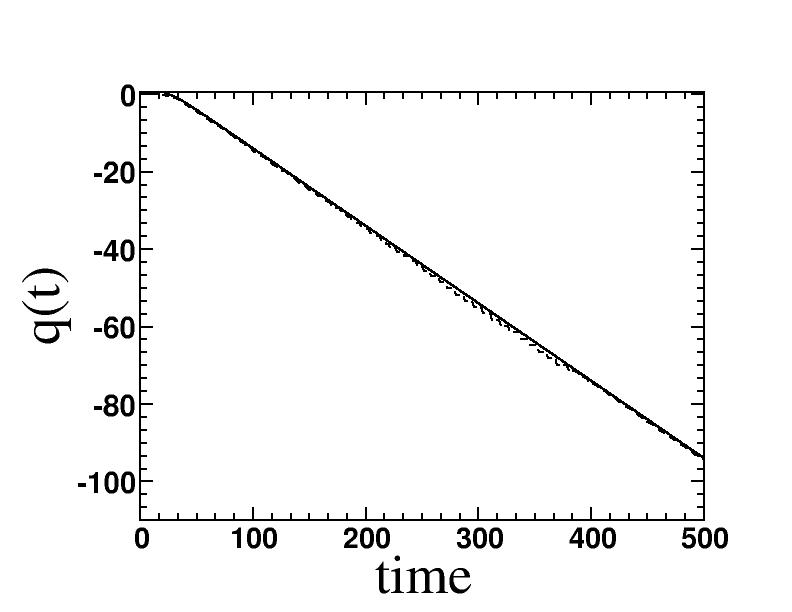} \\
\ & \\
\ & \\
\includegraphics[width=7.0cm]{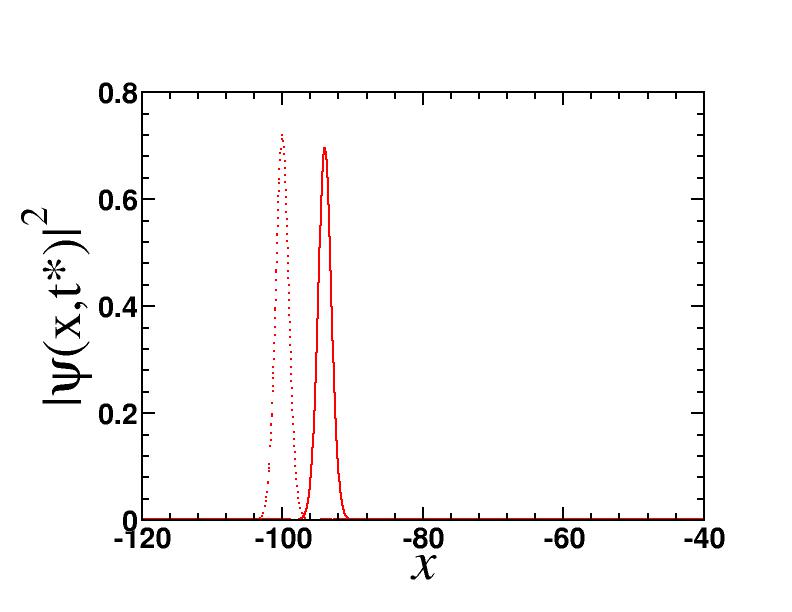}  & 
\quad \includegraphics[width=7.0cm]{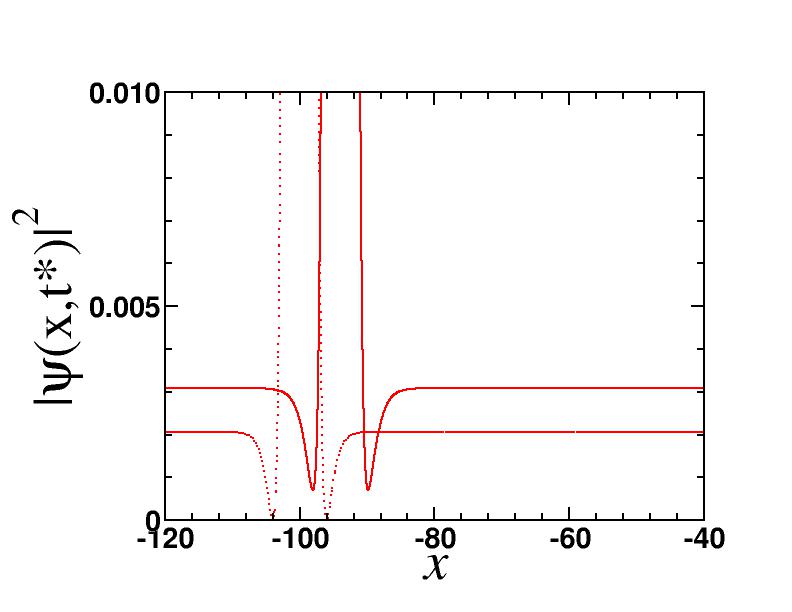}
\end{tabular}
\end{center}
\caption{Comparison of simulations of the collective coordinate (CC) 
approximation and the numerical solution of the PDE where we have considered 
the case $\kappa=1/2$ and included damping when appropriate. Left upper panel: Soliton moving to the left for $t^{*}=250, 500$. 
Right upper panel: Real (solid line) and imaginary (dashed line) 
parts of $\psi(x,t)$ for $t^{*}=500$.  
Real (red solid line) and imaginary (blue 
dashed line) parts of the exact solution for the undamped NLSE for $t=500$. 
Left and right middle panels: comparison of the 
time evolution of $\beta$ and $q$ 
computed from the simulations of the  PDE (solid line) and numerical solutions of the 
CC equations (dashed line).
Lower panels (different scales are shown): for $t^{*}=500$ 
soliton profile from simulations (red solid line) 
and from the exact solution (\ref{psi1}) and (\ref{re1}) (red dashed line).  
Notice that the exact solution is obtained for $\alpha=0$. 
Parameters: $g=2$, $k=0.1$, $r=0.05$, $\delta=-1$, 
$\theta=0$, $\alpha=0.05$, with initial conditions   $\beta_{0}=1$, $p_{0}=0$, $q_{0}=0$ and   
$\phi_{0}=-1.69+\pi/2\approx -0.119$. Notice that for this set of parameters 
the condition $|r|<k'^{4}/(4g)$ (see below Eq. (\ref{re1})) 
is satisfied. After a transient time, since the numerical solution approaches  the 
stationary solution, the $p(v)$ curve is just a point $(-0.1,\,-0.2)$. 
The phase portrait is also 
represented by a point $(0.04491169724,\,-0.02819002364)$. }
\label{fig1} 
\end{figure}


\begin{figure}[ht!]
\begin{center}
\begin{tabular}{cc}
\ & \\
\includegraphics[width=7.0cm]{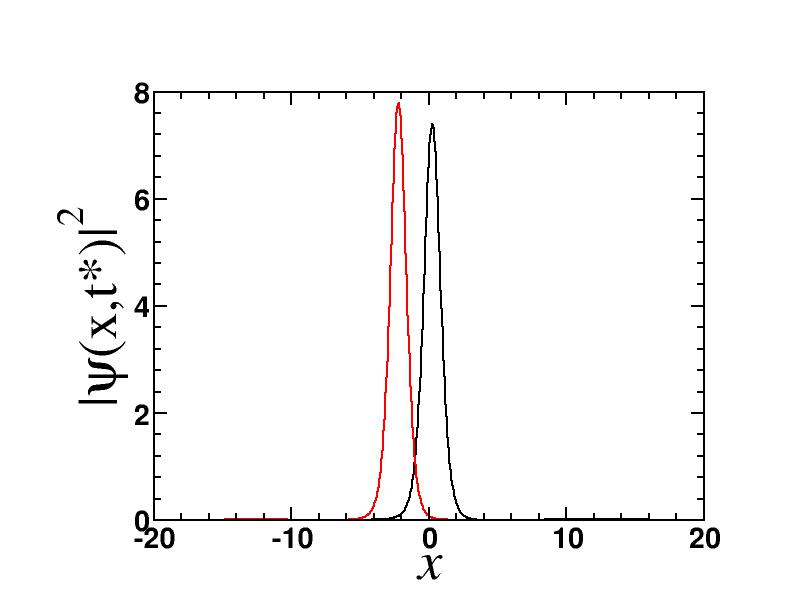}  & 
\quad \includegraphics[width=7.0cm]{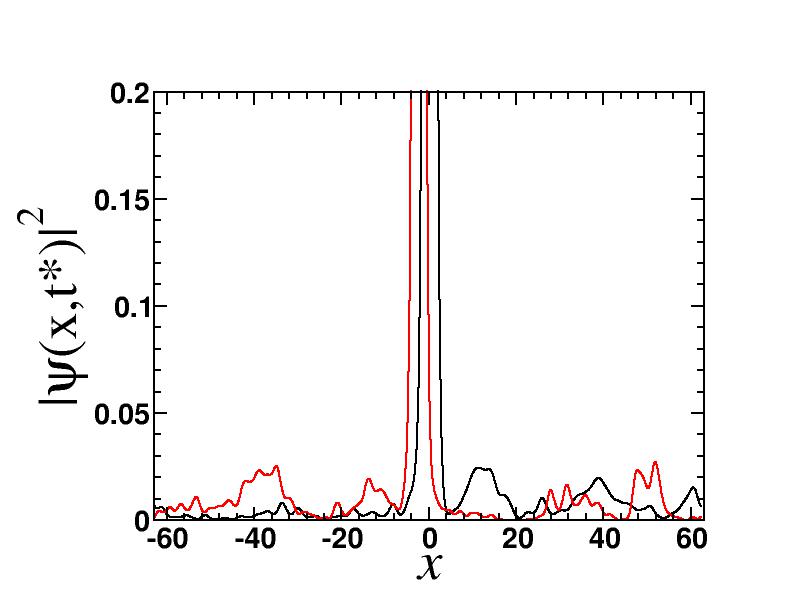}\\
\ & \\
\ & \\
\includegraphics[width=7.0cm]{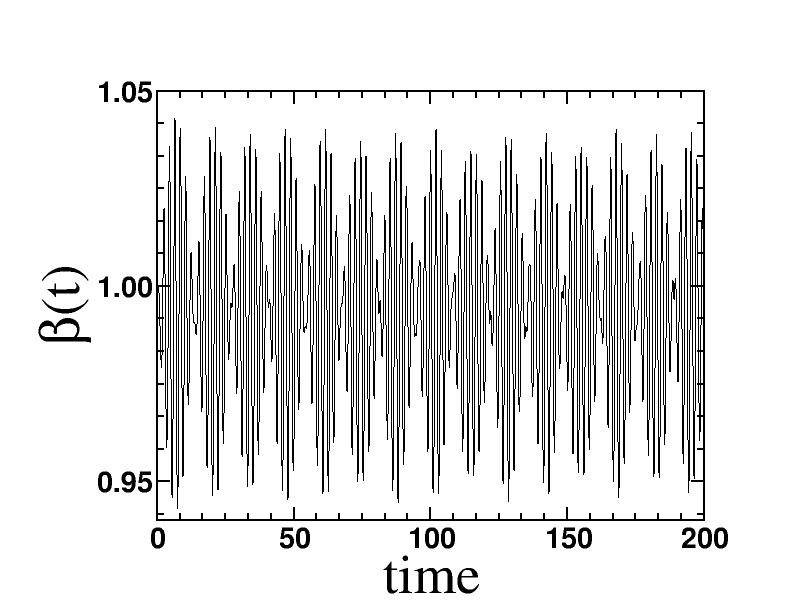}  & 
\quad \includegraphics[width=7.0cm]{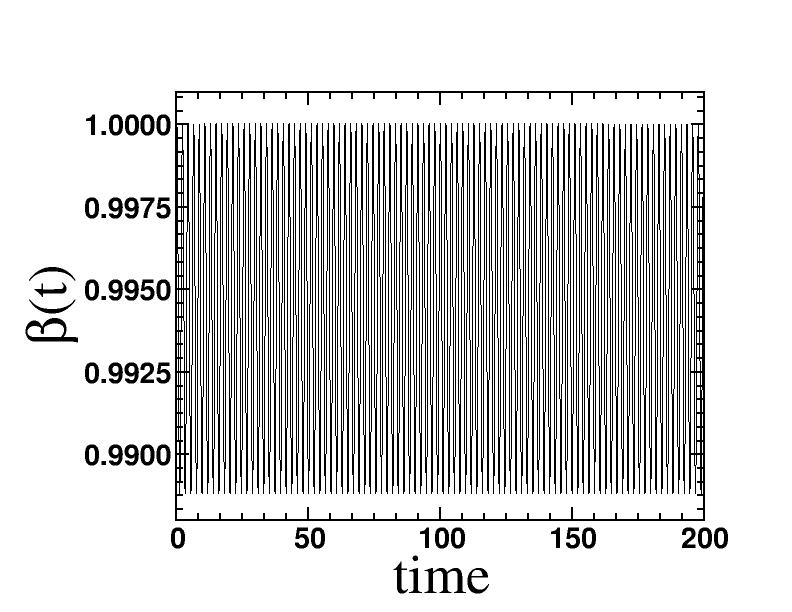}\\
\ & \\
\ & \\
\includegraphics[width=7.cm]{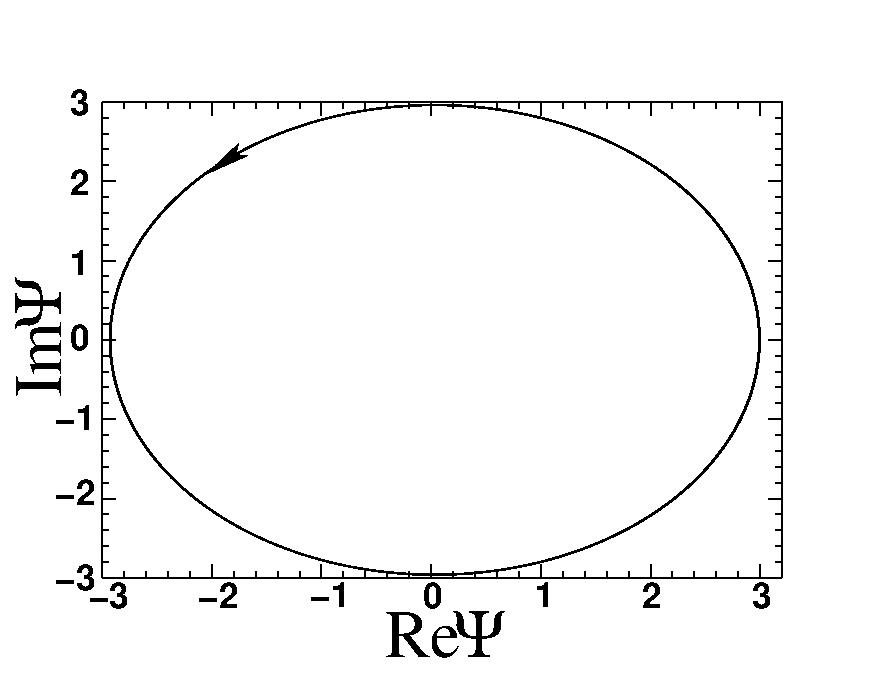} &
\quad \includegraphics[width=7.5cm]{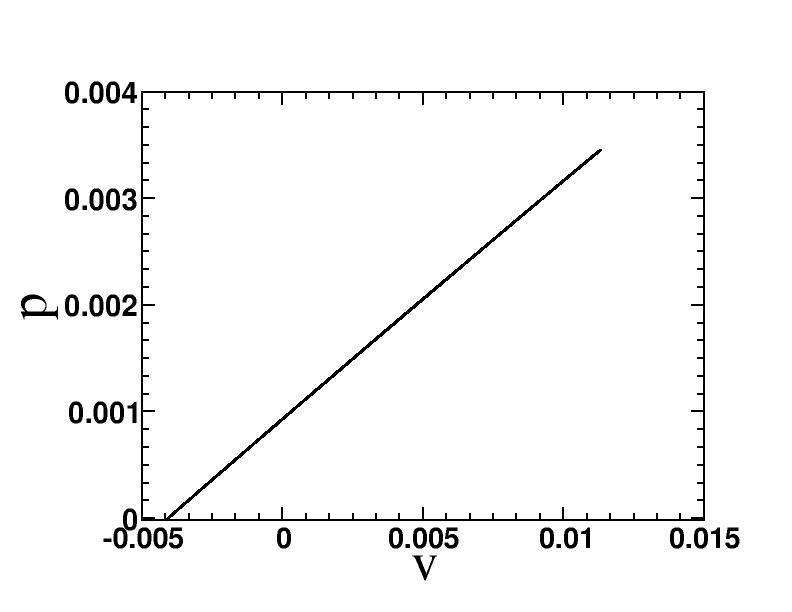} 
\end{tabular}
\end{center}
\caption{Results from simulating the time evolution of the solitary wave 
using both the PDE as well as the CC equations for 
$\kappa=1/2$ with no damping. We use the
same parameters as in Fig \ref{fig1}, but $\alpha=0$.  
Upper panels: soliton profiles (different scales) 
for $t=250$ (black) and $t=500$ (red). 
Middle panels: time evolution of $\beta$ 
computed from the simulations of PDE (left) and 
computed from the numerical solutions of CC equations (right). 
Lower left panel: elliptic orbit in the phase portrait with
positive sense of rotation  which predicts stability;  
lower right panel:  positive slope of the $p(v)$ curve predicts stability.}
\label{fig7} 
\end{figure}

\vspace*{0.5cm}  
\begin{figure}[ht!]
\begin{center}
\begin{tabular}{cc}
 \ & \\
 \ & \\
 \includegraphics[width=6.0cm]{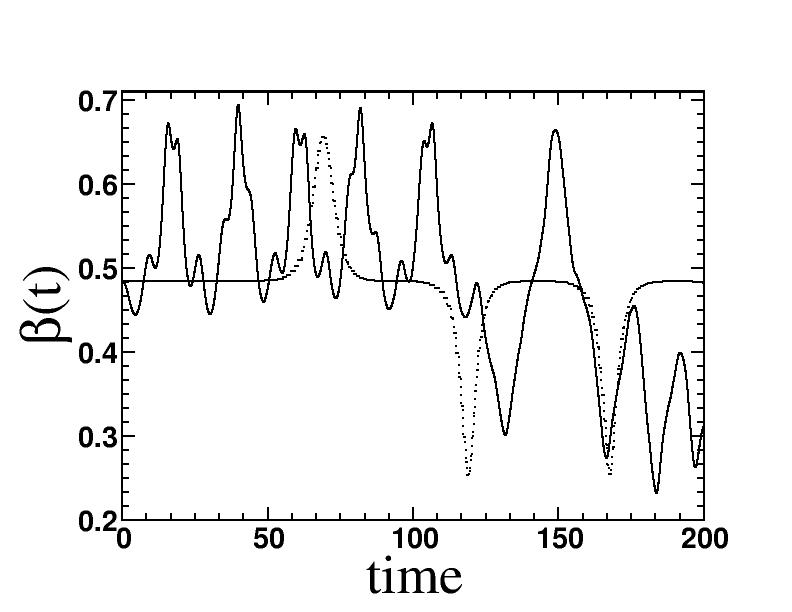}  & 
 \quad \includegraphics[width=6.0cm]{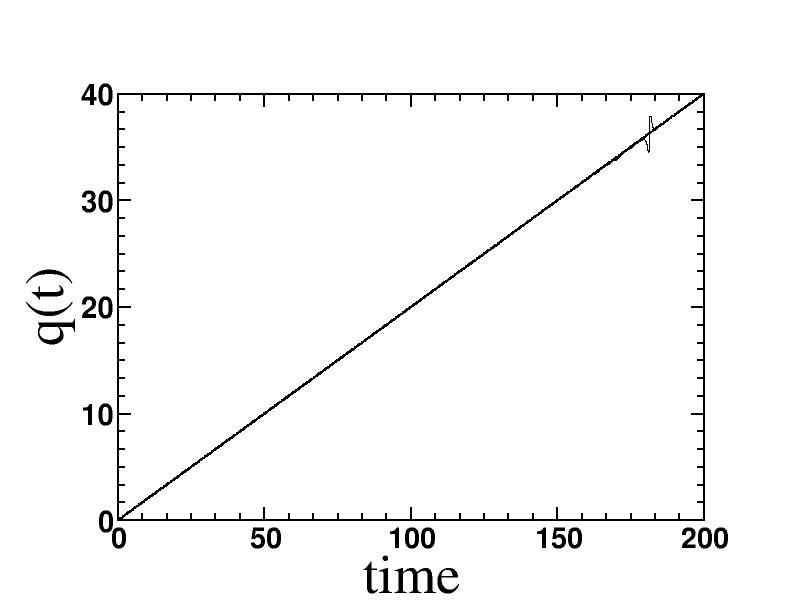} 
\\
 \ & \\
 \ & \\
 \includegraphics[width=6.0cm]{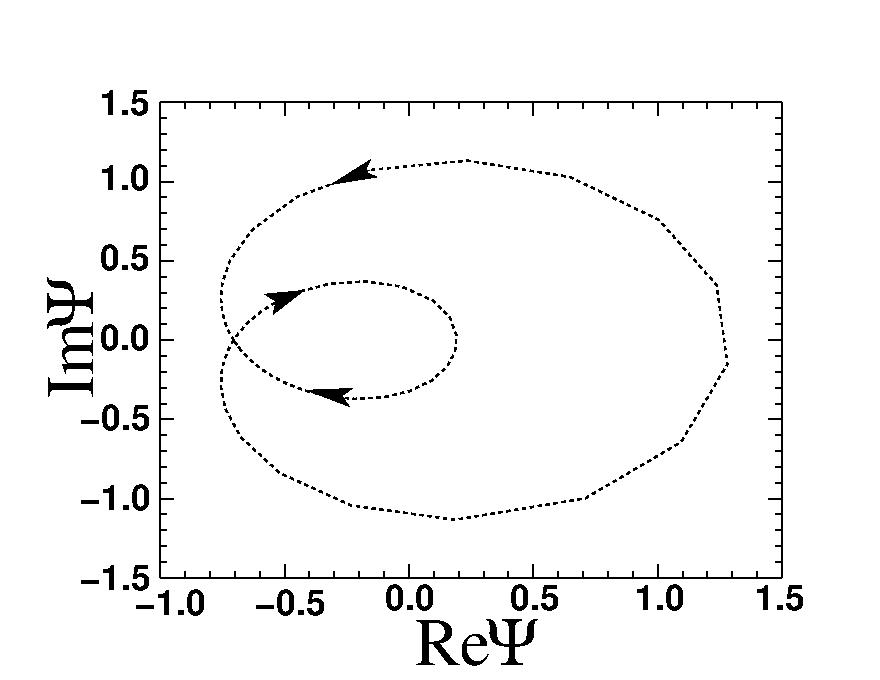}  & 
 \quad \includegraphics[width=6.0cm]{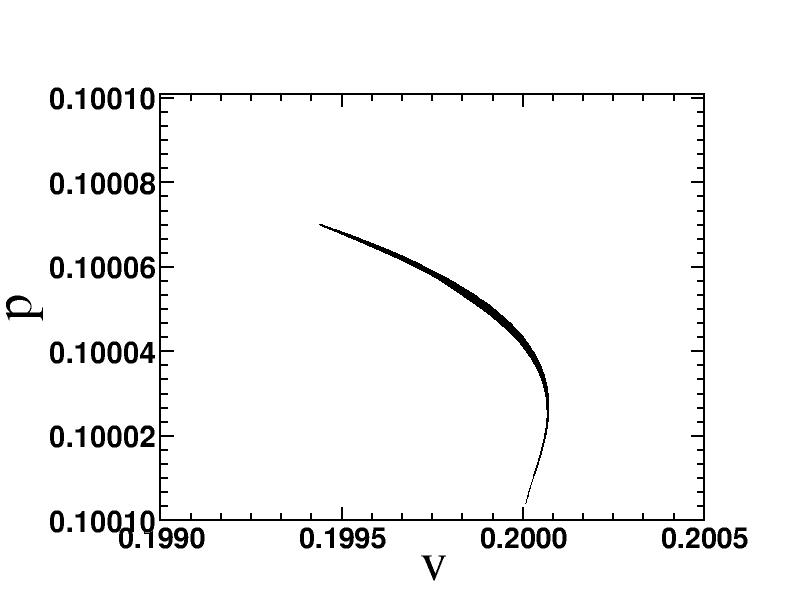} 
\end{tabular}
\end{center}
\caption{Unstable stationary solution for $r=0.05$, $\kappa=1/2$, $g=2$, 
$k=-0.1$, $\theta=0$, $\delta=-1$ and $\alpha=0$. The stationary solution has  $p_0= -k$ (simulations) and 
$p_0= -k+10^{-5}$ (numerical solutions of the CCs), $\phi_0 = \pi$ and 
$\beta_0 =\beta_s=((k'^2+\sqrt{k'^4-8rg/3})/8)^{1/2}$  (see Eq. \ref{f2}). Here we find for $\beta(t)$ that  both 
the simulation (solid line) and the solution to the CC equations (dotted line) 
show switching to a new solution.  From the $p(v)$ curve the negative slope of 
the curve predicts this instability.  In addition, the orbit of the phase portrait is a 
separatrix, which also predicts the instability seen in $\beta$. 
}
\label{fig19A} 
\end{figure}
 \begin{figure}[ht!]
 \begin{center}
 \begin{tabular}{cc}
 \ & \\
 \ & \\
 \includegraphics[width=7.0cm]{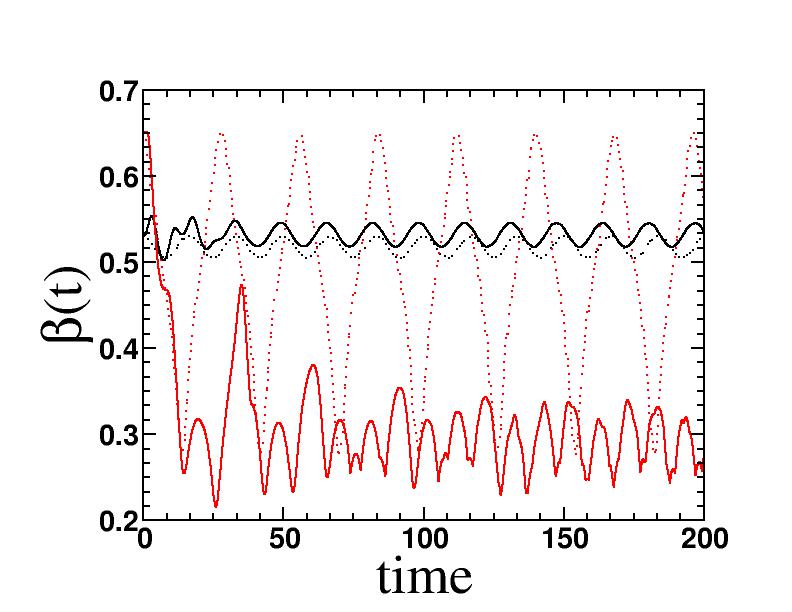}  & 
 \quad \includegraphics[width=7.0cm]{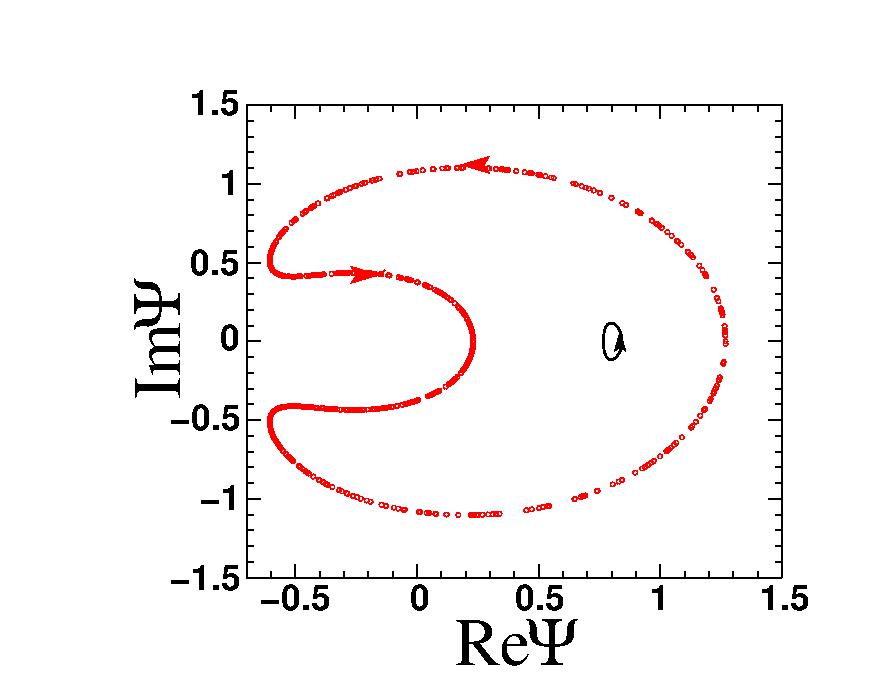} 
 \\
 \ & \\
 \ & \\
 \includegraphics[width=6.0cm]{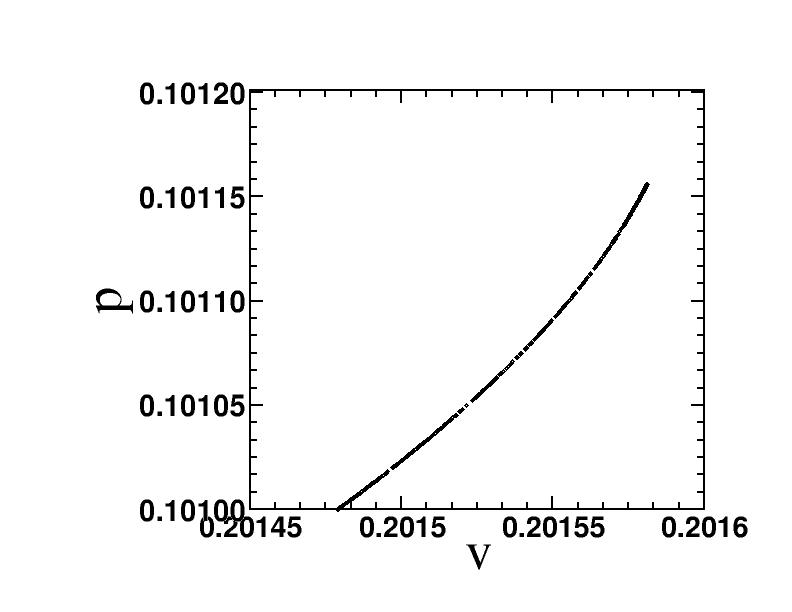}  & 
\quad \includegraphics[width=6.0cm]{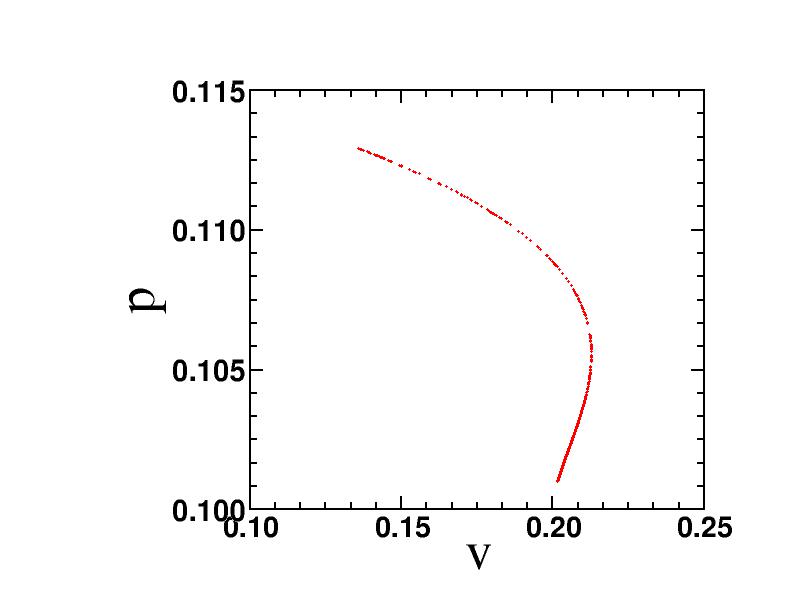} 
 \end{tabular}
 \end{center}
 \caption{Simulations  for the NLSE with $r=0.05$, $k=-0.1$, 
periodic BC, with initial conditions 
$\beta_0 = 0.53$ (black curves) and $0.65$ (red). 
The other parameters and initial conditions are 
$\kappa=1/2$, $g=2$, 
$\delta = -1$, $\theta=0$, $\alpha=0$, $\phi_0=0$, $p_0 = -k+0.001$ and 
$q_0=0$. 
 Upper left panel: $\beta(t)$ compared with numerical solutions of the CC equations (dotted lines). 
Upper right panel: orbits of the phase portrait. 
Lower panels: $p(v)$ curves. 
} 
 \label{fig9pbc} 
 \end{figure}

 \begin{figure}[ht!]
 \begin{center}
 \begin{tabular}{cc}
 \ & \\
 \includegraphics[width=7.0cm]{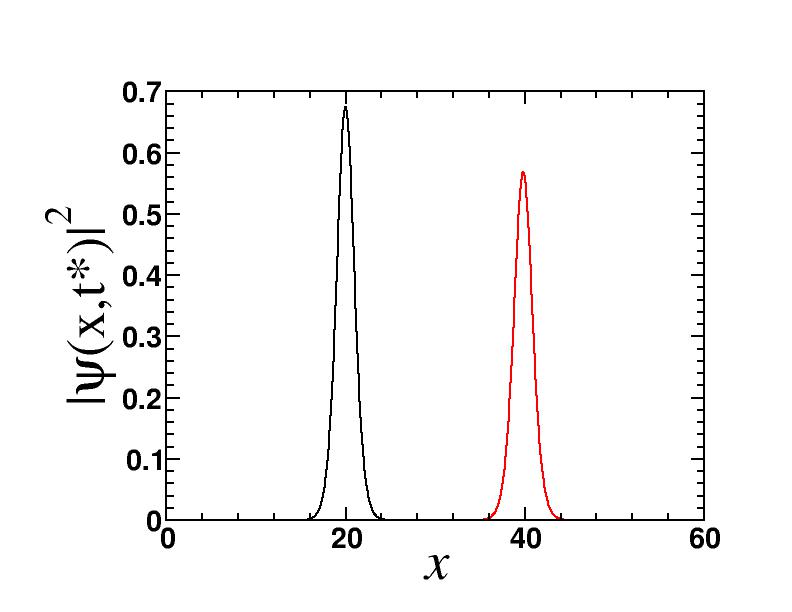} & 
\includegraphics[width=7.0cm]{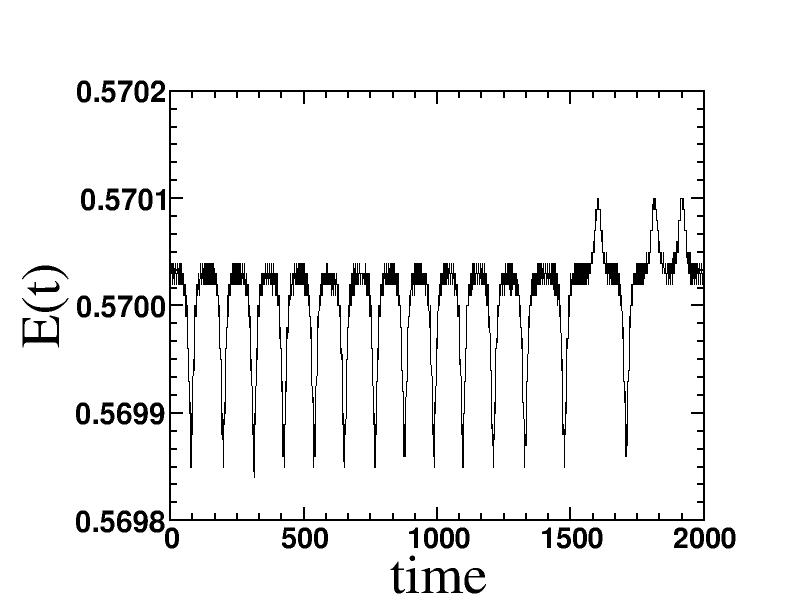}   
 \\
 \ & \\
 \ & \\ 
\includegraphics[width=7.0cm]{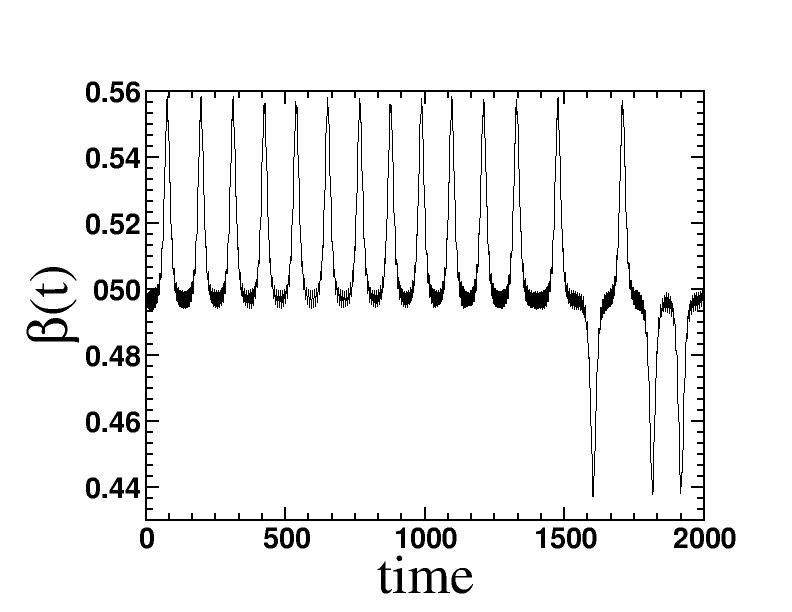} &
 \includegraphics[width=7.0cm]{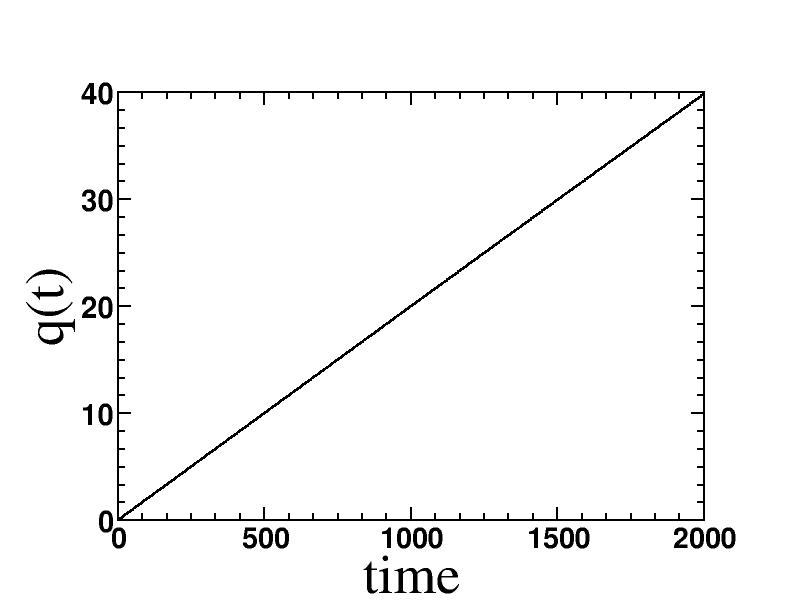}   
 
 \end{tabular}
 \end{center}
 \caption{Numerical simulation of the NLSE for $r=0.05$ using periodic boundary conditions and  $\kappa=1/2$. 
$\beta(t)$ and $E(t)$ show intermittency. 
 Upper panels: Soliton moving to the right for $t^{*}=1000$ (black) and $t^*=2000$ (red) and 
the evolution of the energy of the system. 
 Lower panels: time evolution of $\beta$ and $q$ 
 computed from the simulations of PDE.
 Other parameters of the simulations: $g=2$, $k=-0.01$, $\delta=-1$, 
 $\theta=0$, $\alpha=0$, with initial conditions $p_{0}=p_s=-k$, $q_{0}=0$, $\phi_0=\pi$   
 and $\beta_{0}=\beta_{s}=0.4983$.   This corresponds to the unstable 
stationary solution of Eq.\ (\ref{f2005}), namely  $f_{2+}(y) = 0.745042 \sech^2(0.498345 y)$.
}
 \label{fig12} 
 \end{figure}

\section{Conclusions}
In this paper we have studied analytically, numerically and in a variational approximation the FNLSE with arbitrary nonlinearity parameter $\kappa$.  We studied in detail  a variational approximation based on the solutions to the unforced problem and have studied the CC equations coming from the Euler-Lagrange equations. We determined the stationary solutions of the CC equations and studied their linear stability. 
  We  found that  for small forcing parameter, the CC equations give quantitative agreement with directly solving the FNLSE.  Also the domains of stability of the initial conditions for the FNLSE were quite remarkably close to those found by studying the orbits in the  phase portrait for the CC equations  as well as the stability curves $p(v)$.  We also found that the linearly stable  stationary solutions  of the CC equations were quite close to the stationary solutions of the 
FNLSE, and when these stationary solutions were used as initial conditions for the FNLSE the oscillations of the width parameter were of small amplitude. These simulations at $\kappa = 1/2$ reinforce our belief that our  two dynamical  criteria for understanding the stability of the solitary wave, namely the stability curve $p(v)$ as well as studying orbits in the phase portrait are accurate indications of the stability of the solitary waves as obtained by numerical simulations of the FNLSE.
 We intend to extend our study of the FNLSE equation to the regime $\kappa \geq 2$ to see how forcing affects the blowup of solitary wave solutions in that regime.  This will be done in the variational approximation by adding a variational parameter canonically conjugate to the width parameter $\beta$ similar to the approach of   \cite{stable1} \cite{cooper3}.  Our results are likely 
 to shed light on the behavior of a number of physical systems varying from optical fibers 
 \cite{doped}, Bragg gratings \cite{Bragg}, BECs \cite{bec1, bec2}, nonlinear optics and photonic 
 crystals \cite{crystals}.

\acknowledgments
This work was supported in part by the U.S. Department of Energy. 
F.G.M. acknowledges the hospitality of the Mathematical Institute of the 
University of Seville (IMUS) 
and of the Theoretical Division and Center for Nonlinear Studies at the 
Los Alamos National Laboratory and financial supports by
the Plan Propio of the University of Seville and by Junta de Andaluc\'{\i}a.  
N.R.Q. acknowledges financial support
by the MICINN through FIS2011-24540, and by Junta de Andaluc\'{\i}a
under the projects FQM207, FQM-00481,  P06-FQM-01735 and P09-FQM-4643.


\end{document}